# Pair Wavefunction Symmetry in UTe$_2$ from Zero-Energy Surface State Visualization

Qiangqiang Gu[1]*[§], Shuqiu Wang[1,2,3][§], Joseph P. Carroll[1,4][§], Kuanysh Zhussupbekov[1,4][§], Christopher Broyles[5], Sheng Ran[5], Nicholas P. Butch[6,7], Shanta Saha[6,8], Johnpierre Paglione[6,8], Xiaolong Liu[1,9,10], J.C. Séamus Davis[1,2,4,11]* and Dung-Hai Lee[12,13]*

1. LASSP, Department of Physics, Cornell University, Ithaca, NY 14850, USA
2. Clarendon Laboratory, University of Oxford, Oxford, OX1 3PU, UK
3. H. H. Wills Physics Laboratory, University of Bristol, Bristol, BS8 1TL, UK
4. Department of Physics, University College Cork, Cork T12 R5C, IE
5. Department of Physics, Washington University. St Louis, MO 63130, USA
6. Maryland Quantum Materials Center, University of Maryland, College Park, MD 20742, USA
7. NIST Center for Neutron Research, 100 Bureau Drive, Gaithersburg, MD 20899, USA
8. Canadian Institute for Advanced Research, Toronto, Ontario M5G 1Z8, Canada
9. Department of Physics and Astronomy, University of Notre Dame, Notre Dame, IN 46556, USA
10. Stavropoulos Center for Complex Quantum Matter, University of Notre Dame, Notre Dame, IN, USA
11. Max-Planck Institute for Chemical Physics of Solids, D-01187 Dresden, DE
12. Department of Physics, University of California, Berkeley CA, 94720, USA
13. Materials Sciences Division, Lawrence Berkeley National Laboratory, Berkeley, CA 94720, USA
§    These authors contributed equally to this project.

*Corresponding authors: Qiangqiang Gu, qiangqianggu2016@gmail.com;  J.C. Séamus Davis, jcseamusdavis@gmail.com; Dung-Hai Lee, dunghai@berkeley.edu.

*ABSTRACT*  Although nodal spin-triplet topological superconductivity appears probable in UTe$_2$, its superconductive order-parameter $\Delta_k$ remains unestablished. In theory, a distinctive identifier would be the existence of a superconductive topological surface band (TSB), which could facilitate zero-energy Andreev tunneling to an *s*-wave superconductor, and also distinguish a chiral from non-chiral $\Delta_k$ via enhanced s-wave proximity. Here we employ *s*-wave superconductive scan-tips and detect intense zero-energy Andreev conductance at the UTe$_2$ (0-11) termination  surface. Imaging reveals sub-gap quasiparticle scattering interference signatures with *a*-axis orientation. The observed zero-energy Andreev peak splitting with enhanced s-wave proximity, signifies that $\Delta_k$ of UTe$_2$ is a non-chiral state:  B$_{1u}$, B$_{2u}$ or B$_{3u}$. However, if the quasiparticle scattering along the *a*-axis is internodal, then a non-chiral B$_{3u}$ state is the most consistent for UTe$_2$.

The internal symmetry of electron-pair wavefunctions in non-trivial superconductors (1) is represented by the momentum $\boldsymbol{p} = \hbar\boldsymbol{k}$ dependence of the electron-pairing order parameter $\Delta_{\boldsymbol{k}}$, where $\hbar$ is the reduced Planck constant. For spin-triplet superconductors, where electron-pairs have three spin-1 eigenstates ( $|\uparrow\uparrow\rangle$, $|\downarrow\downarrow\rangle$, $|\uparrow\downarrow+\downarrow\uparrow\rangle$ ), $\Delta_{\boldsymbol{k}}$ is a 2×2 matrix: $\Delta_{\boldsymbol{k}} = \begin{pmatrix} \Delta_{\boldsymbol{k}\uparrow\uparrow} & \Delta_{\boldsymbol{k}\uparrow\downarrow} \\ \Delta_{\boldsymbol{k}\downarrow\uparrow} & \Delta_{\boldsymbol{k}\downarrow\downarrow} \end{pmatrix}$ with $\Delta^{\mathrm{T}}_{-\boldsymbol{k}} = -\Delta_{\boldsymbol{k}}$ and $\Delta_{\boldsymbol{k}} = \Delta^{\mathrm{T}}_{\boldsymbol{k}}$ (1-5). This may also be represented in the $\boldsymbol{d}$-vector notation as $\Delta_{\boldsymbol{k}} \equiv \Delta_0(\boldsymbol{d} \cdot \boldsymbol{\sigma})i\sigma_2$ where $\sigma_i$ are the Pauli matrices. Many such systems should be intrinsic topological superconductors (ITS), where a bulk superconducting energy gap with non-trivial topology co-exists with symmetry-protected TSB of Bogoliubov quasiparticles within that energy gap. Unlike proximitized topological insulators or semiconductors, when three-dimensional (3D) superconductors are topological (6) it is not because of electronic band-structure topology but because $\Delta_{\boldsymbol{k}}$ exhibits topologically non-trivial properties (7). The prototypical example would be a 3D spin-triplet nodal superconductor (1-6) and the search for such ITS which are also technologically viable is a forefront of quantum matter research (8).

Three-dimensional spin-triplet superconductors are complex states of quantum matter (1,4,5). Thus, for pedagogical purposes, we describe a nodal spin-triplet superconductor using a spherical Fermi surface within a cubic 3D Brillouin zone (Fig. 1A). The zeros of $\Delta_{\boldsymbol{k}}$ are then represented by red points at $\pm\boldsymbol{k}_n$. The Bogoliubov-de Gennes (BdG) Hamiltonian is given by:

$$H = \sum_{k_x} \sum_{\boldsymbol{k}_\perp} \psi^+(k_x, \boldsymbol{k}_\perp) \, h(k_x, \boldsymbol{k}_\perp) \psi(k_x, \boldsymbol{k}_\perp). \tag{1}$$

Here $\psi^T(\boldsymbol{k}) = (c_{\boldsymbol{k}\uparrow}, c_{\boldsymbol{k}\downarrow}, c^+_{-\boldsymbol{k}\uparrow}, c^+_{-\boldsymbol{k}\downarrow})$ and $h(k_x, \boldsymbol{k}_\perp)$ is a $4 \times 4$ matrix, containing both band structure and $\Delta_{\boldsymbol{k}}$. We distinguish $\boldsymbol{k} = (k_x, \boldsymbol{k}_\perp)$ because they play different roles in the following didactic presentation. Considering one particular 2D slice of the 3D Brillouin zone with a fixed $k_x$ : its Hamiltonian $h(k_x, \boldsymbol{k}_\perp)$ is that of a 2D superconductor within a 2D Brillouin zone spanned by $\boldsymbol{k}_\perp$. The 2D states $|k_x| < |k_n|$ (blue Fig. 1A) are topological and those $|k_x| > |k_n|$ (green Fig. 1A) are non-topological. The essential signature of such physics is a superconductive TSB (or Andreev bound state (ABS) (7)), on the edges of each 2D slice for $|k_x| < |k_n|$, and its absence when $|k_x| > |k_n|$. The 2D Brillouin zone of any crystal surface parallel to the nodal axis of $\Delta_{\bm{k}}$ is shown in Fig. 1B along with the quasiparticle dispersion $\bm{k}(E)$ of a single TSB. The equatorial circle in Fig. 1B is the $k_x - k_y$ contour satisfying $\epsilon(k_x, k_y, 0) = 0$ with $\epsilon(k)$ being the quasiparticle band dispersion. A line of zero-energy TSB states then connects the two projections of the nodal wavevectors $\pm k_n$ onto this 2D zone (this is often called a "Fermi-arc" although it is actually a two-fold degenerate Majorana-arc of charge-neutral Bogoliubov quasiparticles). Calculation of the density of such TSB quasiparticle states $N(E)$ from $\bm{k}(E)$ in Fig. 1B yields a continuum in the range $-\Delta_0 \leq E \leq \Delta_0$, with a sharp central peak at $E = 0$ due to this arc (Fig. 1C). Thus, 3D nodal spin-triplet superconductors should exhibit a TSB on any surface parallel to their nodal-axis and such TSBs exhibit a zero-energy peak in $N(E)$ (Section 1 of (11)). The conceptual phenomena presented in Figs. 1, A-C, depend solely on whether the symmetry protecting the TSB is broken, and not on material details. Hence, the presence or absence of a gapless TSB on a given surface of a 3D superconductor, of a zero-energy peak in $N(E)$ from its Majorana-arcs, and of the response of the TSB to breaking specific symmetries, can reveal the symmetry and topology of $\Delta_{\bm{k}}$.

UTe$_2$ is now the leading candidate 3D nodal spin-triplet superconductor (9,10). Its crystal symmetry point-group is D$_{2h}$ and the space-group is *Immm* (Section 2 of (11)). Associated with the three basis vectors $\bm{a}, \bm{b}, \bm{c}$ are the three orthogonal $\bm{k}$-space axes $k_x, k_y, k_z$. Within D$_{2h}$ there are four possible odd-parity order parameter symmetries designated A$_u$, B$_{1u}$, B$_{2u}$ and B$_{3u}$ (Section 2 of (11)). All preserve time-reversal symmetry: A$_u$ is fully gapped whereas B$_{1u}$, B$_{2u}$ and B$_{3u}$ have zeros (point nodes) in $\Delta_{\bm{k}}$, whose axial alignment is along $\bm{c}, \bm{b}$ or $\bm{a}$ respectively (Section 2 of (11)). Linear combinations of A$_u$, B$_{1u}$, B$_{2u}$ and B$_{3u}$ are also possible, which break point-group and time-reversal symmetries resulting in a chiral TSB (7,8). For UTe$_2$, there are two chiral states of particular interest with $\Delta_{\bm{k}}$ nodes aligned with the crystal $\bm{c}$-axis, and two with nodes aligned with the $\bm{a}$-axis (Section 2 of (11)). Although identifying which (if any) of these superconductive states exists in UTe$_2$ is key to its fundamental physics, this objective has proven extraordinarily difficult to achieve (12).

Identifying the $\Delta_k$ symmetry of UTe₂ using macroscopic experiments has been problematic because, depending on the sample preparation method, the UTe₂ samples appear to have various degrees of heterogeneity. Samples grown by chemical vapor transport (CV) exhibit small residual resistivity ratios (RRR) (~35) and transition temperatures $T_c \approx 1.6$~2 K (13-15), whereas samples grown by the molten flux method (MF) have larger RRR (~1000) and higher transition temperatures $T_c \approx 2$ K (16). And from macroscopic studies the status $\Delta_k$ for UTe₂ remains indeterminate (17-27) (Section 3 of (11)). To date, $\Delta_k$ symmetry of UTe₂ has been conjectured as non-chiral A$_u$ (17,20), B$_{1u}$ (24), B$_{3u}$ (18,24), chiral A$_u$ + iB$_{3u}$ (21), B$_{2u}$ + iB$_{3u}$ (22), A$_u$ + iB$_{1u}$ (22) and B$_{1u}$ + iB$_{2u}$ (26). Strikingly, however, no tunneling spectroscopic measurements of $\Delta_k$ which could differentiate directly between these scenarios, have been reported.

An efficient tunneling spectroscopic technique for establishing $\Delta_k$ in unconventional superconductors (28-33) is quasiparticle interference imaging (QPI); but this has proven ineffective for unraveling the conundra of UTe₂. This is because conventional single-electron tunneling spectroscopy of UTe₂, even at $T = 280$ mK ($T/T_c \lesssim 1/7$), yields a typical quasiparticle density of states spectrum $N(E \leq \Delta_0)$ that is essentially metallic with only tenuous hints of opening the bulk superconductive energy gap (Fig. 1F) (34,35). Further, UTe₂ surface impedance measurements detect a non-superconductive component of surface conductivity $\sigma_1(\omega, T)$ deep in the superconductive phase (36). Yet the classic QPI signature (37) of a bulk superconductive $\Delta_k$ has been impossible to detect, apparently because the high $N(E \leq \Delta_0)$ overwhelms any tunneling conductance signal from the 3D quasiparticles. Given these challenges to determining the symmetry of $\Delta_k$ using a normal scan-tip, we explored the possibility of using a superconductive scan-tip (38-43 and Section 4 of (11)). Theoretically, we consider two primary channels for conduction from the fully gapped *s*-wave superconductive tip to a nodal spin-triplet superconductor. The first is single-electron tunneling for which the minimum voltage required is $V = \Delta_{\text{tip}}/e$. The second, importantly, is Andreev reflection of pairs of sub-gap quasiparticles (Section 4 of (11)) transferring charge 2*e* across the junction: this occurs because creating or annihilating Cooper pairs costs no energy in a superconductor. Conceptually, therefore, there are notable advantages to using scanned Andreev tunneling spectroscopy for ITS studies, including that TSB quasiparticles within the interface predominate the Andreev process, that the order parameter symmetry difference between sample and tip does not preclude the resulting zero-bias Andreev conductance, and that the enhanced zero-energy conductance peak due to TSB can be detected simply and directly in this way.

To explore this opportunity, we have developed a general guiding theoretical model to describe an *s*-wave superconducting tip (e.g. Nb) connected by tunneling to a nodal *p*-wave superconductor (e.g. UTe$_2$) which sustains a TSB within the interface. We refer to this throughout as the SIP model. To simplify computational complexity, we consider a planar interface shown schematically in Fig. 2A with in-plane momenta as good quantum numbers. The BdG Hamiltonian of this SIP model has three elements: $H = H_{Nb} + H_{UTe_2} + H_T$. Here $H_{Nb}$ is the Hamiltonian for an ordinary *s*-wave superconductor given by $H_{Nb}(k) = \begin{pmatrix} \epsilon_{Nb}(k)\sigma_0 & \Delta_{Nb}(i\sigma_2) \\ \Delta_{Nb}^*(-i\sigma_2) & -\epsilon_{Nb}(-k)\sigma_0 \end{pmatrix}$. Here $\epsilon_{Nb}(k)$ is the band structure model for Nb, $\Delta_{Nb}$ is the Nb superconducting order parameter. $H_{UTe_2}$ is the Hamiltonian of the putative *p*-wave superconductor with $\begin{pmatrix} \epsilon_{UTe_2}(k)\sigma_0 & \Delta_{UTe_2}(k) \\ \Delta_{UTe_2}^+(k) & -\epsilon_{UTe_2}(-k)\sigma_0 \end{pmatrix}$. Here $\epsilon_{UTe_2}(k)$ is the band structure and $\Delta_{UTe_2}(k)$ is a 2 × 2 spin-triplet pairing matrix given by $\Delta_{UTe_2}(k) \equiv \Delta_{UTe_2} i(\mathbf{d} \cdot \boldsymbol{\sigma})\sigma_2$. $H_T$ is the tunneling Hamiltonian between the two superconductors $H_T = -|M|\sum_{k_\parallel}[\psi_{Nb,k_\parallel}^* \sigma_3 \otimes \sigma_0 \psi_{UTe2,k_\parallel}(k) + h.c.]$. Further, $k_\parallel$ is the momentum in the plane parallel to the interface, $\psi$ is the four-component fermion field (Eq. S.2) localizing on the adjacent planes of the *s*-wave and *p*-wave superconductors, while $|M|$ is the tunneling matrix element. To simplify the SIP calculation, $\epsilon_{Nb}(k)$ and $\epsilon_{UTe_2}(k)$ are approximated as single bands (Section 4 of (11)) yet this alters neither the fundamental characteristics of the TSB nor the symmetry properties of the problem, both of which are controlled primarily by the symmetry and topology of $\Delta_k$ (Section 4 of (11)). Finally, our simple band structure model $\epsilon_{UTe_2}(k)$ represents a closed 3D Fermi surface (Section 11 of (11)) upon which depends the non-trivial topology of $\Delta_k$.

For $H_{UTe_2}$ we consider two scenarios: (1) chiral pairing state A$_u$ + iB$_{3u}$ with $\boldsymbol{d(k)} =$ $(0, k_y + ik_z, ik_y + k_z)$ and, (2) non-chiral pairing state B$_{3u}$ with $\boldsymbol{d(k)} = (0, k_z, k_y)$. In both examples the two nodes of $\Delta_{\boldsymbol{k}}$ lie along the $\boldsymbol{a}$-axis as in Fig. 1A, and we use $\Delta_{UTe_2} = \frac{1}{5}\Delta_{Nb}$. First, for $|M| = 0$ we solve the spectrum of $H_{UTe_2}$ exactly. Figure 2B shows the quasiparticle eigenstates $E(k_x = 0, k_y)$ plotted versus $k_y$ for the chiral order parameter with A$_u$ + iB$_{3u}$ symmetry: a chiral TSB spans the full energy range $-\Delta_{UTe_2} \leq E \leq \Delta_{UTe_2}$, crossing the Fermi level ($E$ = 0) and generating a finite density of quasiparticle states $N(|E| < \Delta_{UTe_2})$. Similarly, Fig. 2C shows the quasiparticle spectrum versus $k_y$ at $k_x = 0$ for non-chiral order parameter with B$_{3u}$ symmetry: two non-chiral TSBs span $-\Delta_{UTe_2} \leq E \leq \Delta_{UTe_2}$, and feature $E$ = 0 states generating a finite $N(|E| < \Delta_{UTe_2})$. Although these TSBs have dispersion in both the positive and negative $k_y$ directions and can backscatter, their gaplessness is protected by time-reversal symmetry with $T^2 = -I$. Hence, solely based on $N(|E| < \Delta_{UTe_2})$ of the TSB, one cannot discriminate between the two symmetries of $\Delta_{\boldsymbol{k}}$.

Instead, we explore how to distinguish a chiral from non-chiral $\Delta_{\boldsymbol{k}}$ by using scanned Andreev tunneling microscopy and spectroscopy. Specifically within the SIP model, we calculate the Andreev conductance $a(V) = dI/dV|_{SIP}$ between Nb and UTe$_2$ using the non-chiral TSB and demonstrate that a sharp $a(V)$ peak should occur surrounding zero-bias (Section 7 of (11)). Because the TSB quasiparticles subtending this peak are protected by time-reversal symmetry and because Andreev reflection of TSB quasiparticles allows efficient transfer of charge 2$e$ across the junction, its sharpness is robust. This makes scanned Andreev tunneling spectroscopy an ideal approach for studying superconductive topological surface bands in ITS.

Depending on whether UTe$_2$ is hypothesized as a chiral or non-chiral superconductor, the TSB quasiparticles are themselves chiral (Fig. 2B) or non-chiral (Fig. 2C). As the tunneling matrix element to the $s$-wave electrode $|M| \to 0$ these phenomena are indistinguishable but, as $|M|$ increases, the wavefunctions of the Nb overlap those of UTe$_2$ allowing detection of the TSB quasiparticles at the *s*-wave electrode. Figure 3A shows the predicted quasiparticle bands within the SIP interface for A$_u$ + iB$_{3u}$ symmetry (Fig. 3C) versus increasing $|M|$ (Sections 4 and 5 of (11)). With increasing $|M|\sim 1/R$ where $R$ is the SIP tunnel junction resistance, the proximity effect of the *s*-wave electrode generates two chiral TSBs for all $|E| < \Delta_{\text{UTe}_2}$, both of which cross $E$ = 0. Hence, for the chiral $\Delta_{\boldsymbol{k}}$, the zero-energy $N(E)$ will be virtually unperturbed by increasing $|M|$. Equivalently, Fig. 3B presents the TSB of quasiparticle within the SIP interface as a function of $|M|$ for the non-chiral order parameter with B$_{3u}$ symmetry (Fig. 3C). When $|M| \to 0$ the non-chiral TSB crosses $E$ = 0. But, with increasing $|M|\sim 1/R$, time-reversal symmetry breaking due to the *s*-wave electrode splits the TSB of quasiparticle into two, neither of which cross $E = 0$. This reveals that the $N(0)$ peak must split as the zero-energy quasiparticles of the TSB disappear, generating two particle-hole symmetric $N(E)$ maxima at finite energy. The pivotal concept is thus: whereas the chiral TSB in Fig. 2B requires no symmetry to protect it, the non-chiral TSB of Fig. 2C will open a gap if time-reversal symmetry is broken. This occurs because the SIP model for a non-chiral $\Delta_{\boldsymbol{k}}$ (Fig. 2C) predicts strong $|M|$ locking of the relative phase $\delta\phi$ between the two superconductors at $\delta\phi = \pi/2$ to minimize the total energy of the SIP junction (Sections 4 and 5 of (11)), thus breaking time-reversal symmetry. Contrariwise, the value of $\delta\phi$ is irrelevant for a chiral $\Delta_{\boldsymbol{k}}$ (Fig. 2B) because the TSB at the interface remains gapless for any $\delta\phi$ (i.e., the chiral TSB requires no symmetry to protect it). Figure 3D shows the quantitatively predicted splitting of $N(0)$ into two particle-hole symmetric $N(E)$ maxima as a function of $|M|$ for a chiral $\Delta_{\boldsymbol{k}}$ (orange) and for a non-chiral $\Delta_{\boldsymbol{k}}$ (blue), within the SIP model of Fig. 2A (Sections 4 and 5 of (11)). The decisive fact revealed by this SIP model for Andreev tunneling between an *s*-wave electrode and a *p*-wave topological superconductor through the latter's TSB, is that a non-chiral pairing state can be clearly distinguished from a chiral pairing state.

To search for such phenomena, UTe$_2$ samples are introduced to a superconductive-tip (38-43) scanning tunneling microscope, cleaved at 4.2 K in cryogenic ultrahigh vacuum, inserted to the scan head, and cooled to $T$ = 280 mK. A typical topographic image $T(\boldsymbol{r})$ of the (0-11) cleave surface as measured by a superconductive Nb tip is shown in Ref. 11 Section 8 with atomic periodicities defined by vectors $\boldsymbol{a^*}, \boldsymbol{b^*}$, where $\boldsymbol{a^*}=\boldsymbol{a}$= 4.16 Å is the $\hat{x}$-axis unit-cell vector and $\boldsymbol{b^*}$= 7.62 Å is a vector in the $\hat{y}$:$\hat{z}$ plane. As the temperature is reduced several peaks appear within the overall energy gap: these are clear characteristics of the UTe$_2$ surface states because when the tip is traversed across an adsorbed (non-UTe$_2$) metal cluster the sub-gap peaks disappear (Section 8 of (11)). Most significantly, for Nb scan tips on the atomically homogenous (0-11) UTe$_2$ surface, a sharp zero-energy peak appears in the spectrum as shown in Fig. 4A. This robust zero-bias $dI/dV|_{\text{SIP}}$ peak is observed universally, as exemplified for example by Figs. 4B, C. These phenomena are not due to Josephson tunneling because the zero-bias conductance $a(0)$ of Nb/UTe$_2$ is orders of magnitude larger than it could possibly be due to Josephson currents through the same junction, and because $a(0)$ grows linearly with falling $R$ before diminishing steeply as $R$ is further reduced while $g(0)$ due to Josephson currents should grow continuously as $1/R^2$ (Section 8 of (11)). Moreover, the SIP model predicts quantitatively that such an intense $a(0)$ peak should occur if UTe$_2$ $\Delta_{\boldsymbol{k}}$ supports a TSB within the interface (Fig. 2A), and because Andreev transport due to its quasiparticles allows zero-bias conductance to the Nb electrode (Fig. 2D, Section 7 of (11)).

This discovery provides an exceptional opportunity to explore the TSB quasiparticles of a nodal odd-parity superconductor. To do so we focus on a 44 nm square field of view (FOV) and, for comparison, first image conventional differential conductance at zero-bias $g(\boldsymbol{r}, 0)$ at $T$ = 4.2 K in the normal state of UTe$_2$ as shown in Fig. 4D. The normal-state QPI signature $g(\boldsymbol{q}, 0)$ shown in Fig. 4E, is found from Fourier transform of $g(\boldsymbol{r}, 0)$ in Fig. 4D. Next, Andreev differential conductance $a(\boldsymbol{r}, V) \equiv dI/dV|_{\text{SIP}}(\boldsymbol{r}, V)$ measurements using a superconductive Nb tip are carried out in the identical FOV at $T$ = 280 mK, deep in the UTe$_2$ superconducting state (Fig. 4F and Section 10 of (11)). Note that $a(\boldsymbol{r}, V)$ represents a two-electron process and is thus not proportional trivially to the density of TSB quasiparticle states $N(\boldsymbol{r}, E)$ but, instead, to the Andreev conductance. Our $a(\boldsymbol{r}, 0)$ imaging is then carried out in bias-voltage range $V = 0 \pm 150$ µV inside the $dI/dV|_{\text{SIP}}$ peak (Fig. 4A). Such images introduce atomic-scale visualization of zero-energy quasiparticles of a superconductive TSB.

The Andreev QPI signature $a(\boldsymbol{q}, 0)$ of these zero-energy quasiparticles is shown in Fig. 4G. Here, three new scattering wavevectors $\boldsymbol{S}_{1,2,3}$ are indicated by red circles. Since $\boldsymbol{S}_3$ exists only in the superconducting state and only for $|E| \lesssim 150$ μeV it cannot be due to any new charge ordered state (Section 10 of (11)) but is generated by TSB quasiparticles. And, because a closed Fermi surface has been hypothesized for UTe$_2$ from both angle-resolved photoemission and quantum oscillation research (44,45,46), $\boldsymbol{S}_3$ is not inconsistent with an $\boldsymbol{a}$-axis internodal scattering wavevector on such a Fermi surface.

Finally, to determine spectroscopically whether the UTe$_2$ order parameter is chiral, we measure the evolution of Andreev conductance $a(V)$ at $T$ = 280 mK as a function of decreasing junction resistance $R$ or equivalently increasing tunneling matrix element $|M|$. Figure 5A shows vividly the strong energy splitting $\delta E$ observable in $a(V)$, that first appears and then evolves with increasing $1/R$. Figure 5B shows the measured $a(\boldsymbol{r}, V)$ splitting across the (0 -1 1) surface of UTe$_2$ along the yellow arrow indicated in Fig. 5C for $R$ = 3 MΩ, demonstrating that $a(\boldsymbol{r}, V)$ split-peaks are pervasive. Decisively, from measurements in Fig. 5A, we plot in Fig. 5D the measured $\delta E$ between peaks in $a(\boldsymbol{r}, V)$ at $T$ = 280 mK versus $1/R$. On the basis of predictions for energy splitting $\delta E$ within the SIP model presented in Fig. 3D for chiral $\Delta_{\boldsymbol{k}}$ (Fig. 3A) and non-chiral $\Delta_{\boldsymbol{k}}$ (Fig. 3B), a chiral $\Delta_{\boldsymbol{k}}$ appears ruled out. However, here we note that the SIP model assumes a planar junction with translational invariance parallel to the interface: this implies mirror symmetry ($k_x \to -k_x$) which the STM tip could break, compromising the protection of the non-chiral state and splitting a zero-bias peak (Section 6 of (11)). Nonetheless, since a chiral TSB is symmetry-independent, our conclusion holds: splitting of the zero-bias Andreev conductance peak indicates non-chiral pairing in UTe$_2$.

Thus, the chiral order parameters A$_u$ + iB$_{1u}$ and B$_{3u}$ + iB$_{2u}$ proposed for UTe$_2$ seem inapplicable because of the observed Andreev conductance $a(0)$ splitting (Fig. 5A). Within the four possible odd-parity time-reversal preserving symmetries A$_u$, B$_{1u}$, B$_{2u}$ and B$_{3u}$, the isotropic A$_u$ order parameter appears insupportable because its TSB is a Majorana-cone of Bogoliubons with zero density-of-states at zero energy (7) meaning that Andreev conductance $a(0)$ would be highly suppressed. Among the remaining three possible order parameters B$_{1u}$, B$_{2u}$ and B$_{3u}$, all should exhibit the Andreev conductance $a(0)$ splitting that is observed. However, if the $S_3$ modulations are due to $a$-axis internodal scattering, then the B$_{3u}$ state is favored since its nodes occur along the $a$-axis.

Modeling Andreev conductance from an *s*-wave superconductor through the intervening topological surface band of an intrinsic topological superconductor, reveals a zero-energy Andreev conductance maximum at surfaces parallel to the nodal axis. Further, splitting of this Andreev conductance peak due to proximity of an *s*-wave superconductor signifies a 3D ITS with $\Delta_k$ preserving time-reversal symmetry. Although the B$_{1u}$, B$_{2u}$ or B$_{3u}$ states could all be consistent with such a phenomenology, should the $a(r,0)$ modulations at wavevector $S_3$ result from $a$-axis oriented energy-gap nodes, then the complete experimental data implies that $\Delta_k$ of UTe$_2$ is in the B$_{3u}$ state. Future experiments employing energy-resolved quasiparticle interference imaging of the TSB may explore this premise even more directly. Most generally, use of SIP Andreev conductance spectroscopy for quasiparticle surface band detection and $\Delta_k$ symmetry determination opens new avenues for discovery and exploration of 3D intrinsic topological superconductors.

# FIGURES

**FIG. 1 Pair Wavefunction Symmetry in UTe$_2$**

A. Pedagogical model of a nodal spin-triplet superconductor with order parameter $\Delta_k$ having *a*-axis nodes identified by red dots; the red arrow labels the internodal scattering wavevector. The 2D states $|k_x| < |k_n|$ indicated for example by a blue plane are topological whereas those $|k_x| > |k_n|$ indicated by a green plane are non-topological.

B. The 2D Brillouin zone of the crystal surface parallel to the $\Delta_k$ nodal axis, namely, the *a-b* plane, showing a single TSB dispersion $k(E)$ with color code for *E*. A line of zero-energy TSB states dubbed the Fermi arc connects the two points representing the projections of the 3D $\Delta_k$ nodal wavevectors $\pm k_n(E)$ onto this 2D zone. The equatorial circle in this plot is the $k_x - k_y$ contour satisfies of $\epsilon(k_x, k_y, 0) = 0$ where $\epsilon(k_x, k_y, k_z)$ is the band dispersion used in the model.

C. The density of TSB quasiparticle states $N(E)$ calculated from Fig. 1B exhibits a continuum $|E| \leq \Delta_0$ with a sharp peak at *E* = 0 owing to the TSB Fermi arc.

D. Schematic symmetry of a possible UTe$_2$ order parameter $\Delta_k$ which has two *a*-axis nodes. The *a*-axis oriented internodal scattering $q_n$ is indicated by a red arrow.

E. Schematic of (0 -1 1) cleave surface of UTe$_2$ shown in relative orientation to the STM tip tunneling direction and $\Delta_k$ in Fig. 1D.

F. Measured $N(E)$ of normal (*T* = 4.2 K) and superconducting (*T* = 280 mK) states of UTe$_2$ using a non-superconducting STM tip at the (0 -1 1) cleave surface as seen in Fig. 1E. At the UTe$_2$ surface virtually all states $|E| \leq \Delta_0$ are ungapped.

**FIG. 2 SIP Model: Interfacial Quasiparticle TSB between *p*-wave and *s*-wave Electrodes**

A. Schematic SIP model for interface between an *s*-wave electrode (S) and a *p*-wave superconductor (P) separated by an interface (I), containing the TSB on the surface of the *p*-wave superconductor. There is a variable tunneling matrix element $|M|$ between them, where $|M| \sim 1/R$ and $R$ is the junction resistance. This model is designed to characterize a tunnel junction between superconductive Nb (S) scan-tip and UTe$_2$ surface (P). Any superconductive TSB quasiparticles existing within the interface undergo Andreev scattering between *s*-wave and *p*-wave electrodes.

B. Calculated quasiparticle bands within the SIP interface for a chiral, time-reversal symmetry breaking, *p*-wave order parameter with A$_u$ + iB$_{3u}$ symmetry (Table S2). The Nb electrode has trivial *s*-wave symmetry. For this plot $k_x$ is set to zero. Throughout all the calculated band dispersions, the red dispersion lines denote the superconductive TSB. The shading of the blue dispersion lines is used to highlight the low-energy band structure phenomena, which are central to the tunnelling process within SIP interface.

C. Calculated quasiparticle bands within the SIP interface for a non-chiral, time-reversal symmetry conserving, *p*-wave order parameter with B$_{3u}$ symmetry (Table S1). Here the gapless TSB is protected by time-reversal symmetry. The value of $k_x$ in this plot is set to zero.

D. Schematic of the zero-energy differential Andreev tunneling conductance $a(V) \equiv dI/dV|_{\text{SIP}}$ to the *s*-wave electrode. The magnitude of this zero-bias peak in $a(V)$ is determined by the density $N(0)$ of TSB quasiparticle states within the SIP interface, through a two-quasiparticle Andreev scattering process as shown.

**FIG. 3 Order Parameter Specific TSB Effects with Enhanced Tunneling**

A. Calculated quasiparticle bands within the SIP interface between Nb and UTe$_2$ with $\delta\phi = \pi/2$ as a function of tunneling matrix element $|M|$. Here the chiral order parameter has A$_u$ + iB$_{3u}$ symmetry. As $|M| \to 0, R \to \infty$ the chiral TSB crosses *E* = 0. With increasing $|M|$ (diminishing *R*) the effect of the *s*-wave electrode in the SIP model generates two chiral TSBs inside the UTe$_2$ superconducting gap for all $E < \Delta_{\text{UTe}_2}$, meaning that the zero-energy $dI/dV|_{\text{SIP}}$ peak will be virtually unperturbed (the points where the TSB crossing *E* = 0 are indicated by orange circles).

B. As in Fig. 3A but with a non-chiral TSB which also crosses *E* = 0. With increasing $|M|$ (diminishing *R*) the effect of the *s*-wave electrode splits the quasiparticle bands into two (the split is indicated by blue circles), neither of which crosses *E* = 0. This key observation means that the zero-energy $a(0) = dI/dV|_{\text{SIP}}$ Andreev conductance peak must split into two particle-hole symmetric maxima separating as $|M|$ is increased.

C. Examples of possible order parameter $k$-space phase evolution for UTe$_2$ as used in Figs. 3, A and B. Top panel shows the equatorial ($k_x = 0$) complex phase values of $\Delta_k$ and spin-triplet configurations for chiral order parameter A$_u$ + iB$_{3u}$ (Table S2). Bottom panel shows the equatorial ($k_x = 0$) values of $\Delta_k$ and spin-triplet configurations for non-chiral order parameter B$_{3u}$ (see 11 Table S1). The chiral A$_u$ + iB$_{3u}$ order parameter has a continuous phase winding in contrast to the discontinuous phase change in the B$_{3u}$ order parameter.

D. Calculated energy splitting $\delta E$ of the zero-energy $a(0) = dI/dV|_{\text{SIP}}$ Andreev conductance peak as a function of tunneling matrix element $|M|\sim 1/R$. The $\delta E$ is zero for A$_u$ + iB$_{3u}$ (orange) at all tunneling matrices $|M|$. However, $\delta E$ increases as a function of $|M|\sim 1/R$ for a B$_{3u}$ (blue) order parameter, within the SIP model shown in Fig. 2A. The orange circles correspond to the predicted TSB crossing points in Fig. 3A. The blue circles correspond to the predicted TSB termination points in Fig. 3B.

**FIG. 4 Discovery of Andreev conductance spectrum $a(V)$ for Nb/UTe$_2$ tunneling**

A. Typical SIP Andreev conductance spectrum $a(V) \equiv dI/dV|_{\text{SIP}}$ measured with Nb scan-tip on UTe$_2$ (0 -1 1) surface for junction resistance $R = 6$ MΩ and $T = 280$ mK. A high intensity zero-bias $dI/dV|_{\text{SIP}}$ peak is detected.

B. Typical topographic image $T(r)$ of (0 -1 1) surface ($I_s$ = 0.2 nA, $V_s$ = 5 mV).

C. Evolution of measured $a(r, V)$ across the (0 -1 1) surface of UTe$_2$ indicated by the yellow arrow in Fig. 4B for junction resistance $R = 6$ MΩ and $T = 280$ mK. The zero-bias $dI/dV|_{\text{SIP}}$ peaks are universal and robust, indicating that the zero energy ABS is omnipresent.

D. Measured $g(r, 0)$ at $T = 4.2$ K in the normal state of UTe$_2$.

E. Measured $g(q, 0)$ is the Fourier transform of $g(r, 0)$ in Fig. 4D.

F. Superconductive tip measured $a(r, 0)$ at $T = 280$ mK in the UTe$_2$ superconducting state. This image introduces visualization of the spatial configurations of a zero-energy TSB at the surface of UTe$_2$.

G. Superconductive tip measured $a(q, 0)$ at $T = 280$ mK in UTe$_2$: the Fourier transform of $a(r, 0)$ in Fig. 4F. Three specific new incommensurate scattering wavevectors $S_{1,2,3}$ are indicated by red circles.

**FIG. 5 Evolution and splitting of $a(V)$ peak with enhanced *s*-wave hybridization**

A. Measured evolution of $a(V) \equiv dI/dV|_{\text{SIP}}$ at $T$ = 280 mK in UTe$_2$ as a function of decreasing junction resistance $R$ (i.e. decreasing the tip-sample distance) and thus increasing tunneling matrix element $|M| \sim 1/R$. The $a(V)$ spectra start to split when the junction resistance falls below $R \sim$ 5 MΩ.

B. Evolution of measured $a(\boldsymbol{r}, V)$ splitting across the (0 -1 1) surface of UTe$_2$ along the yellow arrow indicated in Fig. 5C, at junction resistance $R$ = 3 MΩ and $T$ = 280 mK, demonstrating that $a(\boldsymbol{r}, V)$ split-peaks are pervasive at low junction resistance $R$ and high tunneling matrix $|M|$.

C. Topographic image $T(\boldsymbol{r})$ of (0 -1 1) surface ($I_s$ = 0.2 nA, $V_s$ = 3 mV, $T$ = 280 mK) showing the trajectory of the $a(\boldsymbol{r}, V)$ spectra that demonstrate the universality of $a(V)$ splitting in Fig. 5B.

D. Measured energy splitting of $a(V)$ at $T$ = 280 mK in UTe$_2$ versus $1/R$. These data may be compared with predictions of $a(V)$ splitting within the SIP model for A$_u$ + iB$_{3u}$ and B$_{3u}$ order parameters of UTe$_2$ (Fig. 3D).

Figure 1

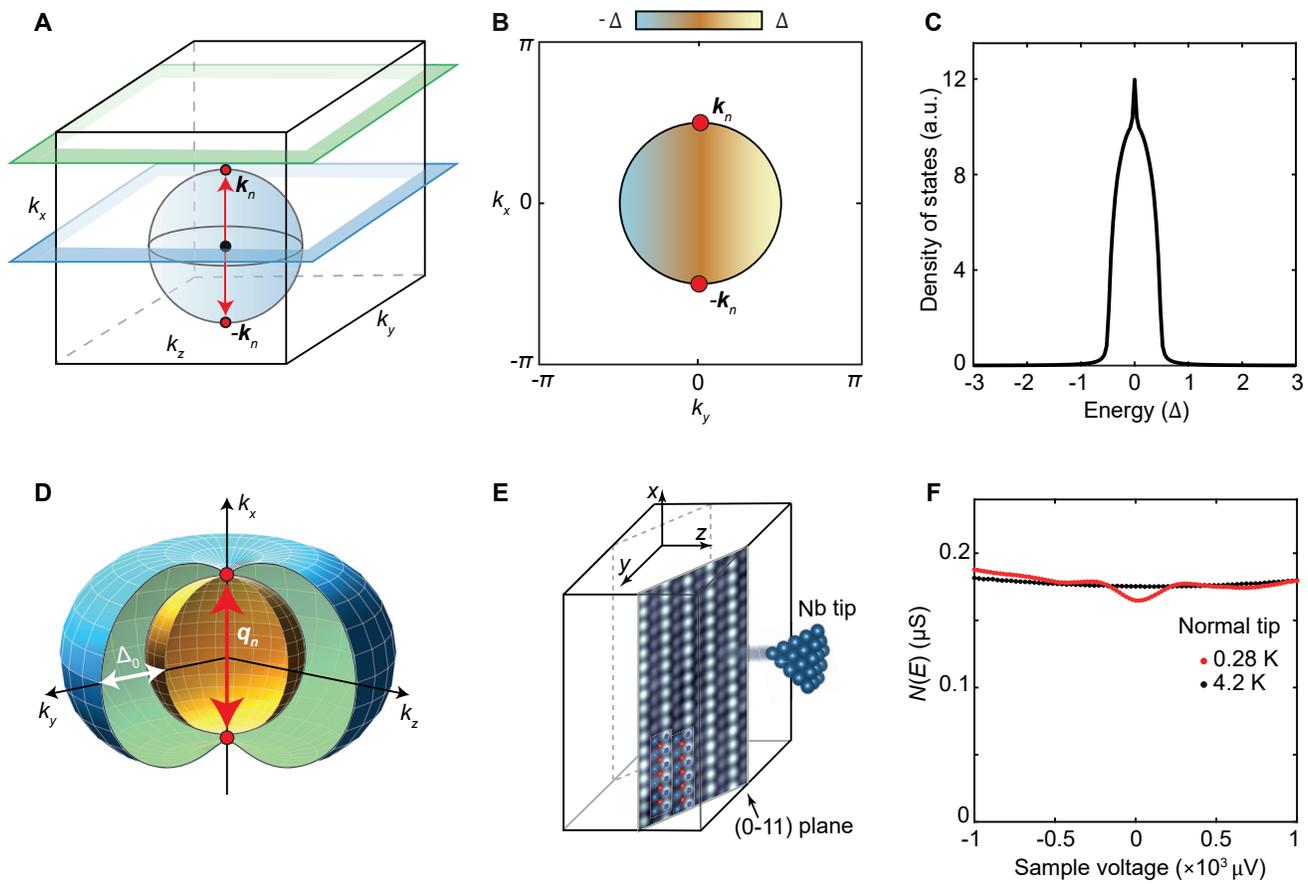

Figure 2

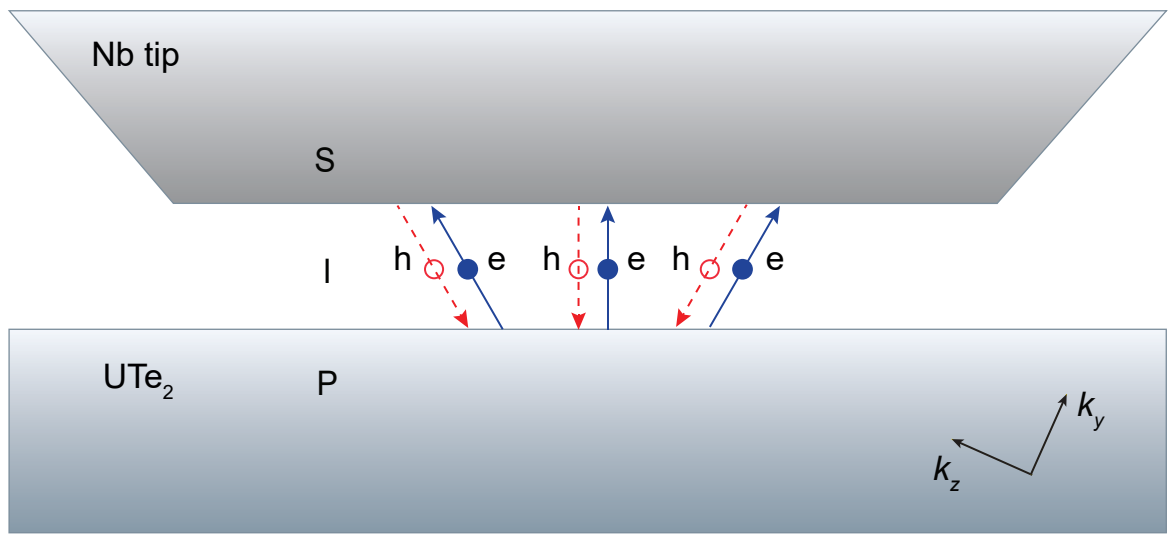
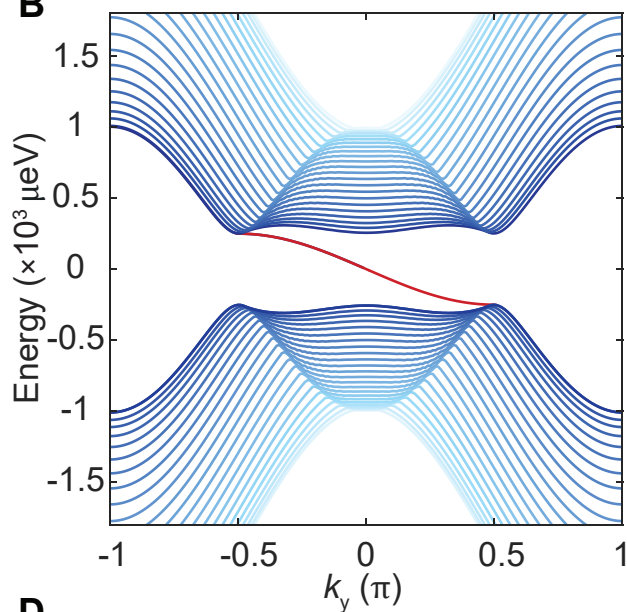
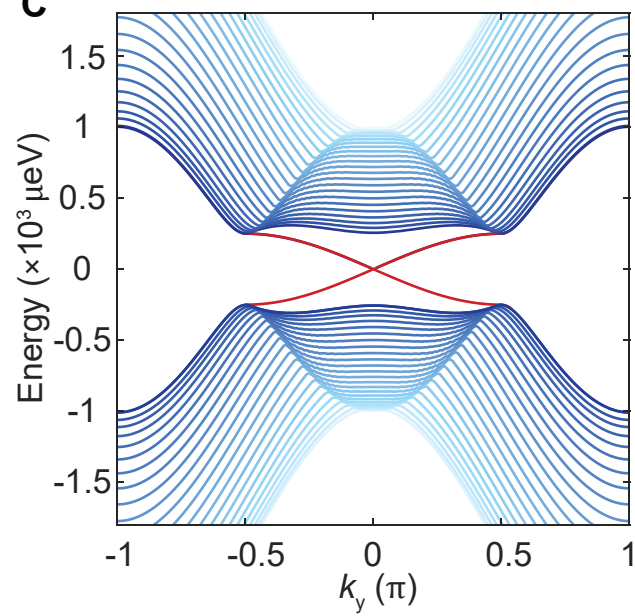
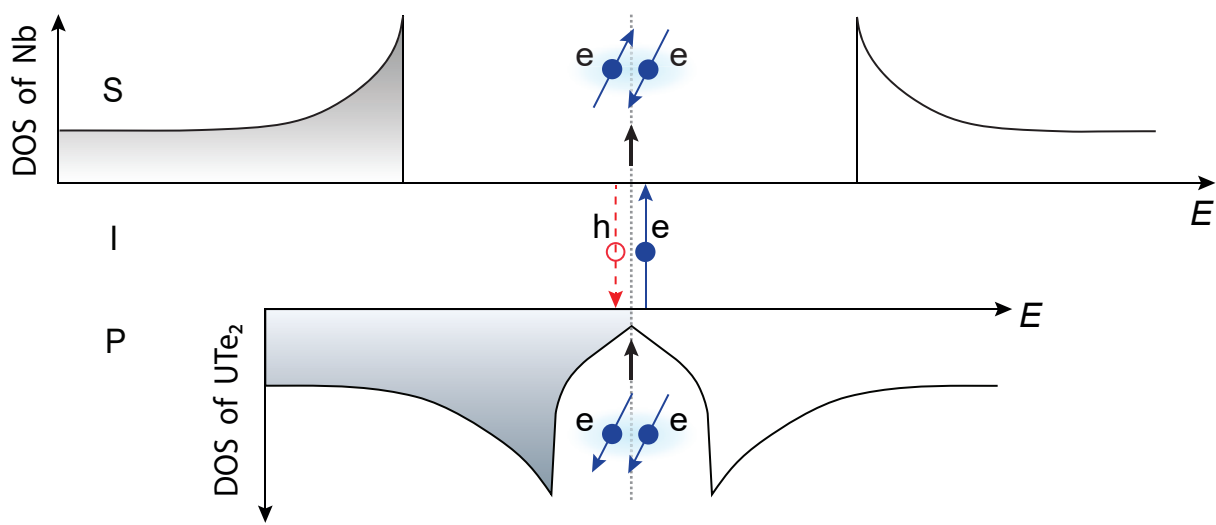

Figure 3

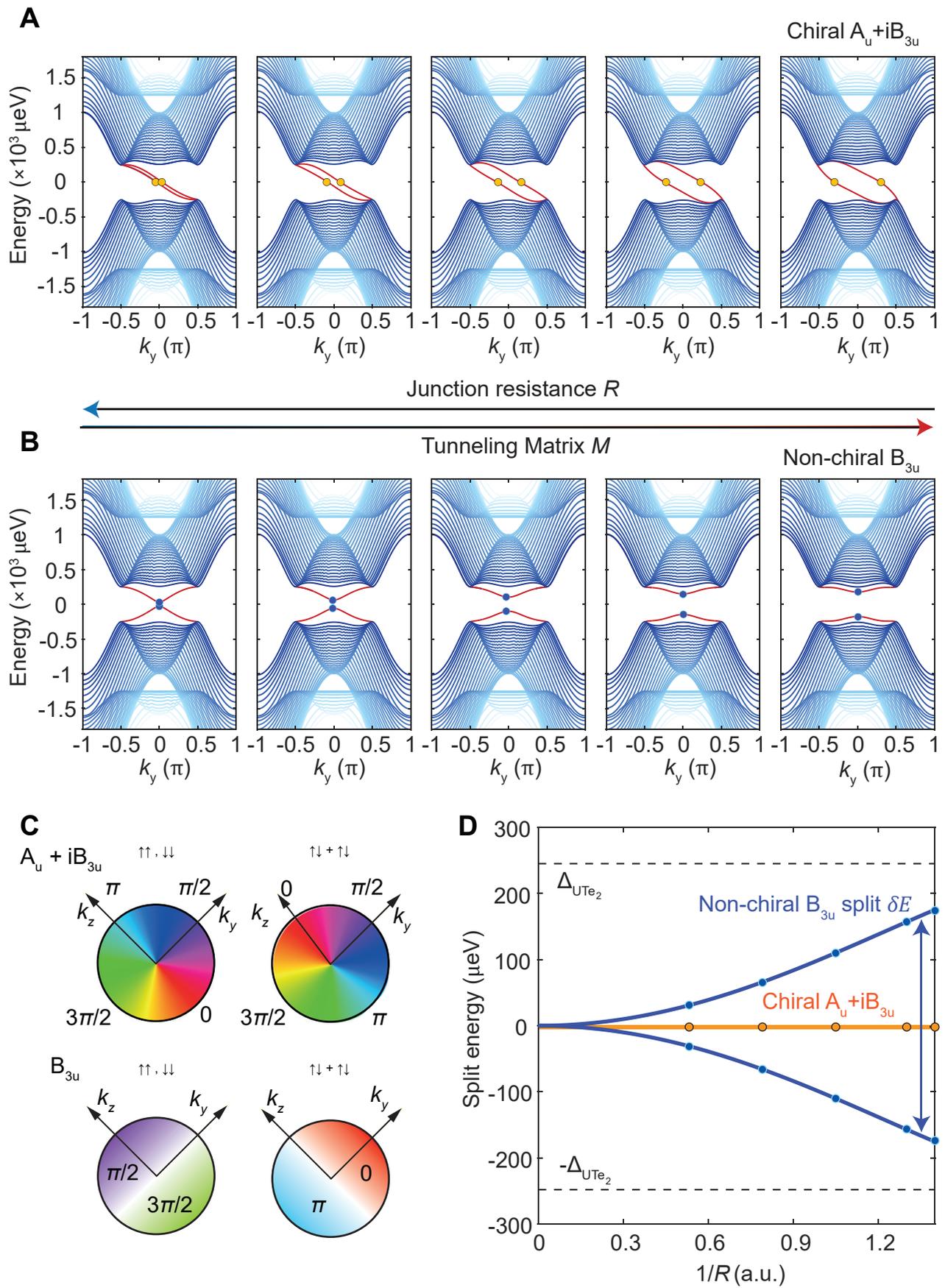

Figure 4

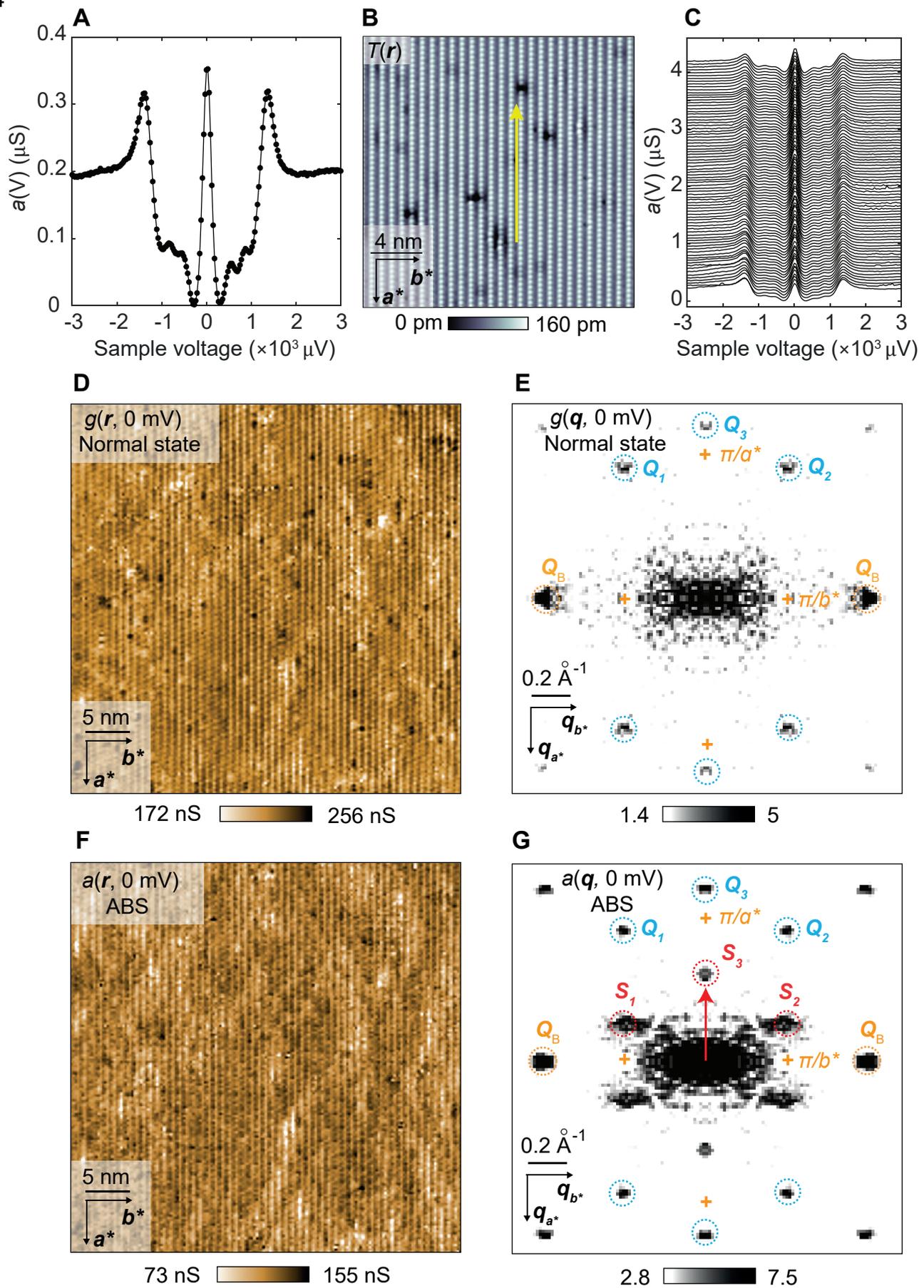

Figure 5

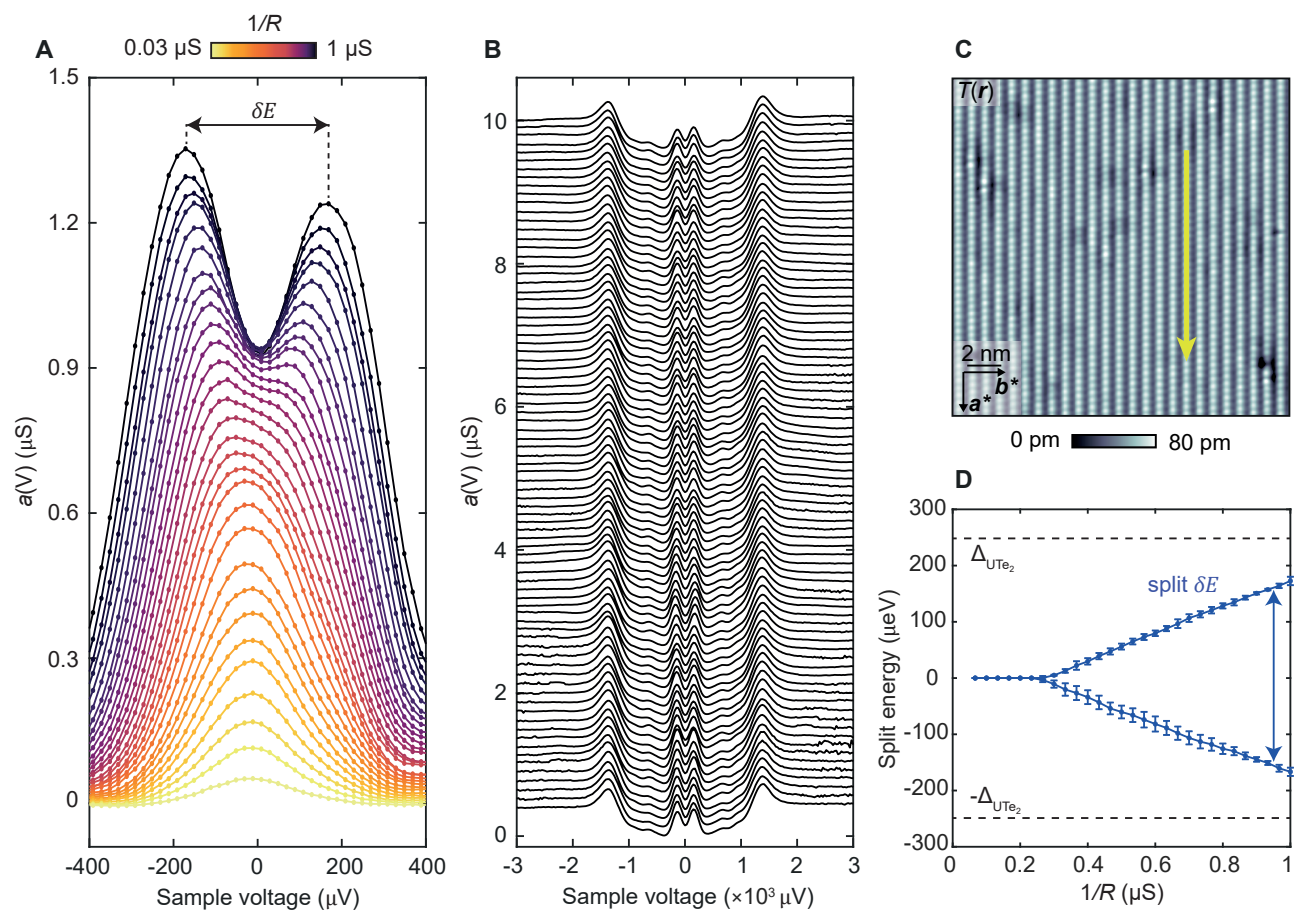

41 X. Liu, Y. X. Chong, R. Sharma, J. C. Davis, Atomic-scale visualization of electronic fluid flow, *Nat. Mater.* **20**, 1480–1484 (2021).

42 S. M. O'Mahony, W. Ren, W. Chen, Y. X. Chong, X. Liu, H. Eisaki, S. Uchida, M. H. Hamidian, J. C. Davis, On the electron pairing mechanism of Copper-Oxide high temperature superconductivity, *Proc. Natl. Acad. Sci. U.S.A.* **119**, e2207449119 (2022).

43 W. Chen, W. Ren, N. Kennedy, M. H. Hamidian, S. Uchida, H. Eisaki, P. D. Johnson, S. M. O'Mahony, J. C. Davis, Identification of a Nematic Pair Density Wave State in $Bi_2Sr_2CaCu_2O_{8+x}$, *Proc. Natl. Acad. Sci. U.S.A.* **119**, 2206481119 (2022).

44 S. Fujimori, I. Kawasaki, Y. Takeda, H. Yamagami, A. Nakamura, Y. Homma, D. Aoki, Electronic structure of $UTe_2$ studied by photoelectron spectroscopy, *J. Phys. Soc. Jpn.* **88**, 103701 (2019).

45 L. Miao, S. Liu, Y. Xu, E. C. Kotta, C.-J. Kang, S. Ran, J. Paglione, G. Kotliar, N. P. Butch, J. D. Denlinger, L. A. Wray, Low Energy Band Structure and Symmetries of $UTe_2$ from Angle Resolved Photoemission Spectroscopy, *Phys. Rev. Lett.* **124**, 076401 (2020).

46 C. Broyles, Z. Rehfuss, H. Siddiquee, K. Zheng, Y. Le, M. Nikolo, D. Graf, J. Singleton, S. Ran, Revealing a 3D Fermi Surface Pocket and Electron-Hole Tunneling in $UTe_2$ with Quantum Oscillations, *Phys. Rev. Lett.* **131**, 036501 (2023).

47 Q. Gu et al., Data associated with 'Pair Wavefunction Symmetry in $UTe_2$ from Zero-Energy Surface State Visualization'. Zendo (2024); https://doi.org/10.5281/zenodo.12676694.

48 D. Agterberg, J. C. Davis, S. D. Edkins, E. Fradkin, D. J. Van Harlingen, S. A. Kivelson, P. A. Lee, L. Radzihovsky, J. M. Tranquada, Y. Wang, The Physics of Pair-Density Waves: Cuprate Superconductors and Beyond, *Annu. Rev. Condens. Matter Phys.* **11**, 231–270 (2020).

49 D. Aoki, H. Sakai, P. Opletal, Y. Tokiwa, J. Ishizuka, Y. Yanase, H. Harima, A. Nakamura, D. Li, Y. Homma, Y. Shimizu, G. Knebel, J. Flouquet, Y. Haga, First Observation of the de Haas–van Alphen Effect and Fermi Surfaces in the Unconventional Superconductor $UTe_2$, *J. Phys. Soc. Jpn.* **91**, 083704 (2022).

50 T. Weinberger, Z. Wu, D. E. Graf, Y. Skourski, A. Cabala, J. Pospisil, J. Prokleska, T. Haidamak, G. Bastien, V. Sechovsky, G. G. Lonzarich, M. Valiska, F. M. Grosche, A. G. Eaton, Quantum

**Acknowledgements:** We acknowledge and thank M. Aprili, C. Bena, J.E. Hoffman, E.-A. Kim, S. Kivelson, A.P. Mackenzie, V. Madhavan and C. Pepin for key discussions and guidance.

**Funding:** Research at the University of Maryland was supported by the Department of Energy Award No. DE-SC-0019154 (sample characterization), the Gordon and Betty Moore Foundation's EPiQS Initiative through Grant No. GBMF9071 (materials synthesis), NIST, and the Maryland Quantum Materials Center. Research at Washington University is supported by the National Science Foundation (NSF) Division of Materials Research Award DMR-2236528. S.W. and J.C.S.D. acknowledge support from the European Research Council (ERC) under Award DLV-788932. X. L. acknowledges support from the Department of Energy (DE-SC0025021). Q.G., S.W., J.P.C. and J.C.S.D. acknowledge support from the Moore Foundation's EPiQS Initiative through Grant GBMF9457. J.C.S.D. acknowledges support from the Royal Society under Award R64897. J.P.C., K. Z. and J.C.S.D. acknowledge support from Science

Foundation Ireland under Award SFI 17/RP/5445. D.-H.L. was supported by the US Department of Energy, Office of Science, Basic Energy Sciences, Materials Sciences and Engineering Division, contract no. DE-AC02-05-CH11231 within the Quantum Materials Program (KC2202).

**Materials and Methods**

Our measurements on UTe₂ crystals were carried out in a custom-built scanned Josephson/Andreev tunneling microscope. The UTe₂ samples were grown by the chemical vapor transport (CV) method as in Ref. 10 and exhibit a $T_c \approx 1.6$ K. The (0-11) surface of the sample was cleaved in cryogenic ultrahigh vacuum at a temperature of ~4.2 K. The sample was then immediately transferred into the STM head. Measurements were carried out using Nb tips at base temperatures of ~4.2 K and ~280 mK. The superconducting Nb tips were prepared by field emission.

**Supplementary Text**

**1. Topological surface bands of nodal spin-triplet superconductors**

Three-dimensional (3D) nodal, odd parity superconductors are analogous to the 3D Weyl semimetal state. As shown in Main Text Fig. 1A, for real-space surfaces parallel to the nodal $\hat{x}$ - axis the in-plane momenta are good quantum numbers. The momentum axis passing perpendicular through the nodes separate the two-dimensional (2D) reciprocal spaces spanned by the in-plane momentum into topologically distinct regions. This is manifested by the presence or absence of topological surface bands (TSBs). In general, the boundary between these topologically inequivalent regions marks the topological phase transition, where the superconducting gap closes.

The 2D Brillouin zone of a crystal surface parallel to the $\Delta_{\boldsymbol{k}}$ nodal axis, namely, the *a-b* plane, shows a line of zero-energy TSB states dubbed a Fermi arc, which connects the two points representing the projections of the 3D $\Delta_{\boldsymbol{k}}$ nodal wavevectors $\pm k_n$ onto this 2D zone. Moreover, the TSB of nodal odd parity superconductors on the surface of *a-b* plane is represented by a 2D band dispersion for $E_{\text{TSB}}(k_x, k_y)$ within the radius of the Fermi surface in the $k_z = 0$ plane (Main Text Fig. 1B). An approximate density-of-states of TSB quasiparticle states can be calculated using

$$N(E) = \sum_{k_{x,y}} \frac{\Gamma/\pi}{\left(E - E_{TSB}(k_x, k_y)\right)^2 + \Gamma^2} \tag{S.1}$$

where $\Gamma$ is a momentum independent quasiparticle broadening parameter (5 μeV is used here). At the surface, when integrated over $(k_x, k_y)$ in the 2D Brillouin zone, the flat Fermi arc at $E = 0$ contributes strongly to a sharp zero-energy peak surface density of states $N(E)$ (Main Text Fig. 1C).

## 2. Candidate superconductive order parameters for UTe$_2$ with D$_{2h}$ symmetry

The crystal symmetry point-group for UTe$_2$ is D$_{2h}$ with space group *Immm* which, if we consider spin-orbit coupling, features four possible, odd-parity order-parameter symmetries: A$_u$, B$_{1u}$, B$_{2u}$, and B$_{3u}$ (Table S1). All single-component representations below preserve time-reversal symmetry and three have zeros (nodes) in $\Delta_k$ whose axial alignment is outlined in Table S1.

| OP | $d$ | $\Delta_k$ | Nodal Axis |
|---|---|---|---|
| A$_u$ | $\alpha k_x \hat{x}$<br>$\beta k_y \hat{y}$<br>$\gamma k_z \hat{z}$ | $(-k_x + ik_y)|\uparrow\uparrow\rangle$<br>$+(k_x + ik_y)|\downarrow\downarrow\rangle$<br>$+k_z(|\uparrow\downarrow\rangle + |\downarrow\uparrow\rangle)$<br>If $\alpha = \beta = \gamma = 1$ | None |
| B$_{1u}$ | $\alpha k_y \hat{x}$<br>$\beta k_x \hat{y}$<br>$\gamma k_x k_y k_z \hat{z}$ | $(-k_y + ik_x)|\uparrow\uparrow\rangle$<br>$+(k_y + ik_x)|\downarrow\downarrow\rangle$<br>If $\alpha = \beta = 1, \gamma = 0$ | $c$ |
| B$_{2u}$ | $\alpha k_z \hat{x}$<br>$\beta k_x k_y k_z \hat{y}$<br>$\gamma k_x \hat{z}$ | $k_z(|\downarrow\downarrow\rangle - |\uparrow\uparrow\rangle)$<br>$+k_x(|\uparrow\downarrow\rangle + |\downarrow\uparrow\rangle)$<br>If $\alpha = \gamma = 1, \beta = 0$ | $b$ |
| B$_{3u}$ | $\alpha k_x k_y k_z \hat{x}$<br>$\beta k_z \hat{y}$<br>$\gamma k_y \hat{z}$ | $ik_z(|\uparrow\uparrow\rangle + |\downarrow\downarrow\rangle)$<br>$+k_y(|\uparrow\downarrow\rangle + |\downarrow\uparrow\rangle)$<br>If $\alpha = 0, \beta = \gamma = 1$ | $a$ |

**Table S1.** Single-component odd-parity spin-triplet superconductive order parameters for D$_{2h}$ symmetry, considering the spin-orbit coupling.

Linear combinations of these D$_{2h}$ order-parameters are also possible which break time-reversal symmetries, resulting in chiral states, shown in Table S2. Two have nodes aligned with the crystal *c*-axis, and two aligned with the *a*-axis. A time-reversal symmetry breaking chiral order parameter breaks down symmetries of the lattice, while remaining an irreducible representation (IR) of the D$_{2h}$ point symmetry group.

| OP | $d$ | $\Delta_k$ | Nodal Axis |
|---|---|---|---|
| $A_u+iB_{1u}$ | $(\alpha_1 k_x + i\alpha_2 k_y)\hat{x}$<br>$(\beta_1 k_y + i\beta_2 k_x)\hat{y}$<br>$(\gamma_1 k_z + i\gamma_2 k_x k_y k_z)\hat{z}$ | $-2k_x\|\uparrow\uparrow\rangle$<br>$+i2k_y\|\downarrow\downarrow\rangle$<br>If $\alpha_{1,2} = \beta_{1,2} = 1\ \gamma_{1,2} = 0$ | c |
| $B_{2u}+iB_{3u}$ | $(\alpha k_z + i\alpha_2 k_x k_y k_z)\hat{x}$<br>$(\beta_1 k_x k_y k_z + i\beta_2 k_z)\hat{y}$<br>$(\gamma_1 k_x + \gamma_2 i k_y)\hat{z}$ | $(k_x + ik_y)(\|\uparrow\downarrow\rangle + \|\downarrow\uparrow\rangle)$<br><br>If $\alpha_{1,2} = \beta_{1,2} = 0\ \gamma_{1,2} = 1$ | c |
| $A_u+iB_{3u}$ | $(\alpha_1 k_x + i\alpha_2 k_x k_y k_z)\hat{x}$<br>$(\beta_1 k_y + i\beta_2 k_z)\hat{y}$<br>$(\gamma_1 k_z + i\gamma_2 k_y)\hat{z}$ | $(ik_y - k_z)(\|\uparrow\uparrow\rangle + \|\downarrow\downarrow\rangle) +$<br>$(k_z + ik_y)(\|\uparrow\downarrow\rangle + \|\downarrow\uparrow\rangle)$<br>If $\alpha_{1,2} = 0, \beta_{1,2} = \gamma_{1,2} = 1$ | a |
| $B_{1u}+iB_{2u}$ | $(\alpha_1 k_y + i\alpha_2 k_z)\hat{x}$<br>$(\beta_1 k_x + i\beta_2 k_x k_y k_z)\hat{y}$<br>$(\gamma_1 k_x k_y k_z + \gamma_2 k_x)\hat{z}$ | $(k_y + ik_z)(-\|\uparrow\uparrow\rangle + \|\downarrow\downarrow\rangle)$<br><br>If $\alpha_{1,2} = 1, \beta_{1,2} = \gamma_{1,2} = 0$ | a |

**Table S2.** Linear combinations of $D_{2h}$ order-parameters give rise to chiral spin-triplet superconductive order parameters with *a*-axis and *c*-axis nodes.

Finally, given the number of free parameters in Tables S1 and S2 there are, of course, an enormous number of other possibilities. For example in the $B_{3u}$ state with $d = (0, \beta k_z, \gamma k_y)$, by choosing real β imaginary γ, their relative phase could (instead of being set at 0 as we do throughout) be chosen as π/2. Another somewhat equivalent example would be a choice of real α imaginary β for the $A_u$ state. However, in these and equivalent cases one is enforcing the breaking of time-reversal symmetry so that the consequent order parameters are not IR of the $D_{2h}$ symmetry group. Furthermore, an admixture of two single components that is non-chiral is allowable but may break further symmetries of the lattice, such as mirror and rotational symmetries. While such order parameters could obviously exist in nature, they are not the subject of our studies nor those of any other UTe$_2$ researchers that we are aware of.

### 3. Experimental evidence of superconductive order parameters

Identifying the $\Delta_k$ symmetry of UTe$_2$ with a specific IR of $D_{2h}$ or with some linear combinations thereof in macroscopic experiments is complicated, and the status of UTe$_2$ $\Delta_k$ remains indeterminate. For example, a magnetic susceptibility upon entering the superconducting phase that is equivalent to Pauli paramagnetism, is deduced from minimal suppressions of the Knight shift (*14,15,17*) and used to adduce spin-triplet pairing (because spin-1 eigenstates typically retain their magnetic moments). Some NMR studies measuring this change of the spin susceptibility across $T_c$ report a decrease in the Knight shift in all directions and hypothesize the isotropically gapped $A_u$ state (*17*), while other NMR studies detect a reduction in the Knight shift along the *b* and *c* axes only, thence hypothesizing $B_{3u}$ pairing symmetry (*18*). Magnetic field orientation of the thermal conductivity (CV) indicates a superconducting energy gap with point nodes parallel to the crystal *a*-axis (*19*), while other field-oriented thermal conductivity (MF) measurements (*20*) report isotropic results and hypothesize an $A_u$ order parameter symmetry. Field-oriented specific heat measurements reveal peaks around the crystal *a*-axis implying point nodes oriented along this direction and hypothesize an order parameter with chiral $A_u + iB_{3u}$ or helical $B_{3u}$ symmetries (*21*). Some electronic specific heat studies (CV) report two specific heat peaks and hypothesize a chiral $A_u + iB_{1u}$ or $B_{2u} + iB_{3u}$ order parameter (*22*), while other specific heat studies (MF) detect only a single specific heat peak and thus hypothesize a single component order parameter (*23*). London penetration depth measurements of superfluid density report anisotropic saturation consistent with nodes along the *a*-axis suggesting $B_{3u}$ symmetry pairing for a cylindrical Fermi surface (*24*), while other penetration depth measurements exhibiting an $n \leq 2$ power law dependence of the penetration depth on temperature motivate a hypothesis of $B_{3u} + iA_u$ pairing symmetry (*25*). Scanning tunneling microscopy experiments (CV) in the (0-11) plane parallel to *a*-axis show energy-reversed particle-hole symmetry breaking at opposite mirror-symmetric $UTe_2$ step edges (*26*) with the consequent hypothesis of a chiral surface state $B_{1u} + iB_{2u}$ whose nodes are aligned to the *a*-axis. Polar Kerr effect measurements (CV) report a field-induced Kerr rotation indicating the presence of time-reversal symmetry breaking and hypothesize chiral $B_{2u} + iB_{3u}$ or $A_u + iB_{1u}$ pairing (*22*) with nodes aligned to the *c*-axis, while other polar Kerr effect measurements (MF) report no detectable spontaneous Kerr rotation (*27*).

## 4. SIP model

We model a planar interface between an *s*-wave and a nodal *p*-wave superconductor (SIP), in which is located a topological surface band. Firstly, we construct the general four component Bogoliubov-de Gennes (BdG) Hamiltonian for a superconductor,

$$H = \sum_k \psi^+(\boldsymbol{k}) H(\boldsymbol{k}) \psi(\boldsymbol{k}), \psi(\boldsymbol{k}) = (c_{\boldsymbol{k}\uparrow}, c_{\boldsymbol{k}\downarrow}, c^+_{-\boldsymbol{k}\uparrow}, c^+_{-\boldsymbol{k}\downarrow})^\mathrm{T} \qquad (S.2)$$

$H_{Nb}$ is the Hamiltonian of an *s*-wave superconductor Nb with

$$H_{Nb}(\boldsymbol{k}) = \begin{pmatrix} \epsilon_{Nb}(\boldsymbol{k})\sigma_0 & \Delta_{Nb}(i\sigma_2) \\ \Delta^*_{Nb}(-i\sigma_2) & -\epsilon_{Nb}(-\boldsymbol{k})\sigma_0 \end{pmatrix} \qquad (S.3)$$

Here $\epsilon_{Nb}(\boldsymbol{k})$ is the band structure of Nb, $\Delta_{Nb}$ is the Nb superconducting order parameter (in the computation we take $\Delta_{Nb}$ to be momentum independent), and $\sigma_{0,1,2,3}$ are the four components of Pauli matrices. $H_{UTe_2}$ is the Hamiltonian of the putative *p*-wave superconductor with

$$H_{UTe_2}(\boldsymbol{k}) = \begin{pmatrix} \epsilon_{UTe_2}(\boldsymbol{k})\sigma_0 & \Delta_{UTe_2}(\boldsymbol{k}) \\ \Delta^+_{UTe_2}(\boldsymbol{k}) & -\epsilon_{UTe_2}(-\boldsymbol{k})\sigma_0 \end{pmatrix} \qquad (S.4)$$

Here $\epsilon_{UTe_2}(\mathbf{k})$ is the band structure containing a model spherical Fermi surface of UTe2, and $\Delta_{UTe_2}(\mathbf{k})$ is a 2 × 2 odd parity pairing matrix given by $\Delta_{UTe_2}(\mathbf{k}) \equiv \Delta_{UTe_2} i(\mathbf{d} \cdot \boldsymbol{\sigma})\sigma_2$. Both $\epsilon_{UTe_2}(\mathbf{k})$ and $\epsilon_{Nb}(\mathbf{k})$ are nearest neighbor tight binding dispersions on a simple cubic lattice with the same lattice constant and are modeled as $\cos(k_x) + \cos(k_y) + \cos(k_z) - 2$ in the unit of meV. Given the focus of this project on the interplay of pairing symmetry, the exact band structures for Nb and UTe2 are irrelevant.

We compare two candidate order parameters for UTe2, one built from the IR B3u of Table S1. This is the non-chiral gap function with nodes in the $\mathbf{a}$ direction. The second order parameter considered is the chiral state Au + iB3u with nodes also in the $\mathbf{a}$ direction (for B1u + iB2u pairing the conclusions are the same). Below we present the 2 × 2 pairing matrix in momentum space derived using $\Delta_k \propto i[\mathbf{d}(\mathbf{k}) \cdot \boldsymbol{\sigma}]\sigma_2$.

For B3u: $\mathbf{d} = (0, k_z, k_y)$

$$\Delta_k \propto \begin{pmatrix} ik_z & k_y \\ k_y & ik_z \end{pmatrix} \tag{S.5}$$

For Au + iB3u: $\mathbf{d} = (0, k_y + ik_z, k_z + ik_y)$

$$\Delta_k \propto \begin{pmatrix} -k_z + ik_y & k_z + ik_y \\ k_z + ik_y & -k_z + ik_y \end{pmatrix} \tag{S.6}$$

Lastly, we model the tunneling Hamiltonian between the Nb and UTe2 interfaces as

$$H_T = -|M| \sum_{\mathbf{k}_\parallel} [\psi^\dagger_{Nb,\mathbf{k}_\parallel} \sigma_3 \otimes \sigma_0 \psi_{UTe_2,\mathbf{k}_\parallel} + h.c.] \tag{S.7}$$

where $\mathbf{k}_\parallel = (k_x, k_y, 0)$ is the quasiparticle momentum parallel to the interface. The model developed consists of twenty Nb layers adjacent to fifty UTe2 layers stacked in the $\hat{\mathbf{c}}$ direction, with tunneling between the surface layer of each superconductor (Fig. S1A). We then derive the eigenvalues and eigenvectors of the total Hamiltonian $H = H_{Nb} + H_{UTe_2} + H_T$. We keep those eigenenergies whose wavefunction weight exceeds a certain lower bound ($10^{-3}$ weight on the top surfaces of Nb and UTe2) and plot the eigenvalues of the Hamiltonian against $k_y$ with $k_x$ as a parameter. For all calculations we set the superconducting gap magnitude of UTe2 to be 0.25 meV to approximate the gap magnitude of the sample, and set the Nb gap to 1.25 meV to approximate the gap of the STM tip.

At $k_x = 0$, when the tunneling matrix $M = 0$, the band dispersion on Nb top layer shows a continuum within the full Nb gap energy (Fig. S1B). Fig. S1C shows the band dispersion on UTe2 top layer, TSBs are formed on the UTe2 surface indicated by red lines and the extra bulk bands are indicated by blue lines. The observables are dominated by the topological surface band consisting of one sheet for the A$_u$ + iB$_{3u}$ model or two sheets for B$_{3u}$ alone.

To highlight TSB states, Figures 2B,C of the main text show the band dispersion on the top surface of UTe$_2$ at $k_x = 0$ for chiral A$_u$ + iB$_{3u}$ and non-chiral B$_{3u}$ state when $M$ is set to zero, without the effect of Nb. We have checked that the nature of the spectrum remains independent of the bound so long as the number of layers is sufficiently large.

For a chiral superconducting order parameter with symmetry A$_u$ + iB$_{3u}$, the TSB has a two-fold degenerate chiral TSB starting from negative $\Delta_{\text{UTe}_2}$, positive $k_y$ to positive $\Delta_{\text{UTe}_2}$, negative $k_y$ (Main Text Fig. 2B). This is the expected chiral surface state dispersion in which the TSB quasiparticles break time-reversal symmetry. In comparison, the non-chiral order parameter with symmetry B$_{3u}$ develops two TSB branches, which are symmetric with respect to the $k_y = 0$ axis and thus do not break time-reversal symmetry (Main Text Fig. 2C). The spectra for both chiral and non-chiral order parameters feature TSBs for $-0.5\pi < k_x < 0.5\pi$ where $(\pm 0.5\pi, 0, 0)$ are the locations of the gap nodes. For $|k_x| > 0.5\pi$ there are no in-gap states and at $k_x = \pm 0.5\pi$ the gap closes. Figures S1E,F show the 3D representation of these two chiral and non-chiral TSB.

To investigate the tunneling effect between Nb and UTe$_2$, we calculate the band dispersion on Nb top surface in Fig. S1D when $|M|$ is nonzero (0.2 meV is used in the calculation). The tunneling process can be categorized into two types based on the energy range. Outside the Nb gap $\Delta_{\text{Nb}}$, the UTe$_2$ bulk states are overlapped with Nb states via single-particle tunneling. Inside $\Delta_{\text{Nb}}$, the bulk states of UTe$_2$ states cannot penetrate into the bulk of Nb and contribute to the tunneling conductance. However, and most importantly, we find that the TSBs of UTe$_2$ can tunnel into the bulk of Nb in the form of Cooper pairs by the Andreev reflection process without the energy cost of the Nb gap. (Fig. S1D).

## 5. Hybridization of a *p*-wave TSB with an *s*-wave electrode

To investigate the effect of tunneling between a Nb tip and UTe$_2$ we set a finite tunneling amplitude $|M| > 0$ and plot the $k_x = 0$ BdG spectrum derived from the three-term Hamiltonian $H = H_{\text{Nb}} + H_{\text{UTe}_2} + H_{\text{T}}$. In Main Text Figs. 3A and B we demonstrate the effect of increasing the magnitude of $|M|$ (which corresponds to decreasing the tunnel junction resistance experimentally). When $|M|$ increases in the SIP model for a chiral A$_u$ + iB$_{3u}$ superconductor, the surface state develops two branches at different momenta while maintaining zero-energy crossings (Main Text Fig. 3A). Thus, the Andreev conductance peak in $dI/dV|_{\text{SIP}}$ remains as a single maximum at zero-energy with reducing STM junction resistance. As $|M|$ increases in the SIP model for a non-chiral B$_{3u}$ superconductor, the TSB splits into two particle-hole symmetric energy bands with symmetric momentum dependence with respect to the $k_y = 0$ axis. Thus, the zero-energy Andreev conductance peak in $dI/dV|_{\text{SIP}}$ must split into two finite energy $dI/dV|_{\text{SIP}}$ maxima which further move apart in energy as the junction resistance is reduced. From the $N(E)$ calculation of the TSB as implemented in Eqn. S1, we find that with increasing $|M|$, the single maximum at zero-energy in $N(E)$ remains unchanged for chiral $A_u + iB_{3u}$ state while it splits into two peaks for non-chiral $B_{3u}$ state.

The key to understanding the TSB splitting is the relative phase $\delta\phi$ between the Nb and UTe$_2$ order parameters at the interface. When $|M| = 0$, the gapless edge states of the B$_{3u}$ pairing superconductor are protected by time-reversal symmetry ($T^2 = -I$ and Z$_2$ classification). When $|M| > 0$ for the B$_{3u}$ state the relative phase $\delta\phi$ evolves to $\frac{\pi}{2}$ (Fig. S2B) lowering the total electronic energy of the system by reducing the energies of all occupied TSB states $E < 0$. When $\delta\phi = \frac{\pi}{2}$ the time-reversal symmetry within the SIP junction is broken because upon the time-reversal $e^{i\delta\phi} = i \rightarrow e^{-i\delta\phi} = -i$. Thus, if UTe$_2$ is an odd-parity superconductor with non-chiral B$_{3u}$ state whose isolated $\Delta_k$ preserves time-reversal symmetry, the SIP relative phase becomes $\delta\phi = \frac{\pi}{2}$ due to proximity of the $s$-wave electrode. Under such condition the gapless TSB is no longer protected. Conversely, if UTe$_2$ is a chiral superconductor with an order parameter such as $A_u + iB_{3u}$ then TRS is broken without the influence of the $s$-wave superconducting STM tip. In that case the topological classification is Z, and the gapless TSB does not require any symmetry to remain protected and so remains gapless regardless of $|M|$ (Fig. S2A).

## 6. The effect of mirror symmetry breaking by the STM tip

Another potential cause of the TSB splitting phenomenology is possible if the gap function of UTe$_2$ exhibits B$_{3u}$ pairing. In that case the gap function is independent of $k_x$. Consequently, the BdG Hamiltonian is invariant under an anti-unitary symmetry ($T_1$): $\hat{T}_1^{-1}\Psi_{k_x,k_y,k_z}T_1 = iI\sigma_y\Psi_{k_x,-k_y,-k_z}$. Furthermore, since the SIP model assumes planar tunneling, where translational symmetry parallel to the surface is preserved, an additional mirror symmetry ($M_r$) exists: $M_r^{-1}\Psi_{k_x,k_y,k_z}M_r = \Psi_{-k_x,k_y,k_z}$. Together, $M_r \cdot T_1$ restores time-reversal symmetry. In an STM setup involving point tunneling, the mirror symmetry ($M_r$) can be broken, which compromises time-reversal symmetry. This symmetry breaking allows a gap to open in the non-chiral TSB. In general, this effect also becomes more pronounced as the tip-sample conductance increases.

## 7. Andreev conductance of $s$-wave electrode through a $p$-wave TSB

A key consideration is the effect of hybridization of a $p$-wave TSB with an $s$-wave electrode on the Andreev conductance across the junction between the $p$-wave and $s$-wave superconductors. The origin of Andreev reflection for superconductors is the anomalous term $\sum_k \sum_{\alpha,\beta}[\Delta_{\alpha\beta}(k)c^+_{\alpha,k}c^+_{\beta,-k} + h.c.]$ (here $\alpha$ and $\beta$ label the spin of the electron) in the Hamiltonian. This term allows an incident electron (hole) impacting on an order parameter $\Delta_{\alpha,\beta}(\mathbf{k})$ to reflect as a hole (electron) as depicted in Main Text Fig. 2A.

Most simply, a single Andreev reflection transfers two electrons (holes) between the tip and the sample (Main Text Fig. 2D). Based on a S-matrix approach, the formula to compute the Andreev conductance of the SIP Model is

$$a(V) = \frac{8\pi^2 t_{\text{eff}}^4 e^2}{h} \sum_n \frac{\langle \phi_n|P_h|\phi_n\rangle\langle \phi_n|P_e|\phi_n\rangle}{(eV - E_n)^2 + \pi^2 t_{\text{eff}}^4 [\langle \phi_n|P_h|\phi_n\rangle + \langle \phi_n|P_e|\phi_n\rangle]^2} \quad (S.8)$$

Here $|\phi_n\rangle$ is the projection of the $n^{\text{th}}$ TSB eigenfunction onto the top UTe$_2$ surface, and $P_e$ and $P_h$ are the electron and hole projection operators acting on the UTe$_2$ surface and $V$ is the bias voltage. Note that the Andreev conductance $a(V)$ is different from $N(E)$ in Eqn. S1. However, both equations count the related eigenvalues through the integral over the whole TSB.

Thus, in principle, the sharp zero-energy peak of the calculated $N(E)$ in Main Text Fig. 1C will be clearly reflected in the sharp zero-energy peak of Andreev conductance $a(V)$. Figure S3A shows the schematic of the TSB generated Andreev reflection to the *s*-wave electrode. Specifically within the SIP model, we plot in Fig. S3B the calculated $a(V)$ from Eqn. S8 which predicts a sharp peak in Andreev conductance surrounding zero-bias. In this figure we have divided the total Andreev conductance by the number of transverse $k_\parallel$ channels to mimic the point tunneling of STM.

## 8. Topological surface band and Andreev phenomenology at UTe$_2$ (0-11) surface

A typical topograph of the UTe$_2$ crystal cleave surface (0 -1 1) is presented in Fig. S4A and its Fourier transform is shown in Fig. S4B. To demonstrate empirically that the zero-energy state detected in $dI/dV|_{\text{SIP}}$ is a result specifically of the UTe$_2$ TSB we present a linecut across a cluster of impurity atoms in Fig. S5. This cluster is likely made up of Nb atoms which have been accidentally transferred from the Nb STM tip to the surface of the sample. The sub-gap conductance quickly collapses to zero and the $dI/dV|_{\text{SIP}}$ spectra become fully gapped as the tip measures across the metallic cluster (Fig. S5). The zero-energy $dI/dV|_{\text{SIP}}$ peak on the unperturbed surfaces is therefore not an artefact of the Nb scanning tip, but an omnipresent feature of the UTe$_2$ (0 -1 1) surface.

Here it is important to emphasize the zero bias peak cannot be attributed to the Josephson current. This can be demonstrated clearly by plotting typical measured Andreev zero-bias conductance $a(0)$, as shown in Fig. S6A, versus tip-sample junction resistance $R$ on the same plot with the maximum possible zero-bias conductance which could be generated by the Josephson effect $g(0)$, here exemplified by measured Nb/NbSe$_2$ Josephson zero-bias conductance data. These results are presented in Fig. S6B. At high $R$, the intensity of measured $a(0)$ of Nb/UTe$_2$ is many orders of magnitude larger than it could possibly be due to Josephson currents. Moreover measured $a(0)$ for Nb/UTe2 first grows linearly in decreasing $R$ but then diminishes steeply as $R$ is reduced further, whereas the zero-bias conductance due to Josephson currents $g(0)$ grows rapidly as $1/R^2$ as exemplified in the Nb/NbSe2 $g(0)$ data. Most importantly one does not expect the zero-bias peak due to the Josephson effect to split when $R$ is low. These highly repeatable and internally consistent experimental facts demonstrate the absence of detectable Josephson currents between Nb electrodes and the UTe2 (0-11) termination surface.

## 9. Phase fluctuation effect on tip-induced time-reversal symmetry breaking

These data and SIP model raise the issue of fluctuations in the relative phase $\delta\phi$ between the Nb and UTe2 order parameters when interacting predominantly by Andreev coupling. Recall that if UTe2 is an odd-parity superconductor with a nodal, non-chiral, time-reversal conserving state $\Delta_k$, the minimum energy SIP relative phase is $\delta\phi = \frac{\pi}{2}$ due to proximity of the *s*-wave electrode. This effect will spit the zero-bias Andreev conductance as shown in Fig. S6A. To evaluate if thermal fluctuations in $\delta\phi$ should wipe out the peak splitting effect for the realistic parameterization of $\Delta_k$ of UTe2, temperature $T$ and junction resistance $R$, we calculate the TSB density-of-states $N(E)$ when $\delta\phi = \frac{\pi}{2}$ and when $N(E)$ is averaged over the whole range $0 < \delta\phi < \pi$. The result as presented in Fig. S7 demonstrates that realistic phase fluctuations will not wipe out *s*-wave tip-induced $N(E)$ splitting, thus preserving the Andreev $a(V)$ conductance splitting.

## 10. Imaging and Fourier analysis of $dI/dV|_{\text{SIP}}$: Andreev conductance modulation $S_3$

Imaging Andreev conductance reveals spatial modulations in the zero-energy $dI/dV|_{\text{SIP}}$ (Fig. S8). These data are measured at a junction resistance of 5 MΩ in the same field of view as the topograph in Fig. S8A. Fourier transformation of this $a(r, 0)$ Andreev conductance map, $a(q, 0)$, shows new features that only exist in the superconducting state. Among them is the new wavevector $S_3$ (Fig. S8) whose identification requires the following considerations.

First, if $S_3$ is a normal-state CDW, it must appear above $T_C$. But in all our experimental studies $S_3$ is only observed in the superconducting phase (Main Text Fig. 4G). Moreover, a CDW is highly unlikely to exist only in the very narrow energy range of ~ ±150 μeV (Fig. S8D) where $S_3$ is observed. Thus, $S_3$ cannot be considered a normal-state CDW.

Second, interaction between uniform superconductivity of UTe2 with a pre-existing CDW (*35*) or PDW (*34*) both occurring with the same wavevector $Q$, cannot induce a CDW or a PDW at $Q/2$ as this is ruled out by Ginzburg-Landau theory (*48*).

Third, in the SIP model the projected gap nodes on the surface BZ are not isolated $k$ points since they are connected by a zero energy Fermi arc leading to finite DOS. Thus, the narrow energy distribution of $S_3$ is more consistent with QPI pertinent to the gap nodes along the $k_x$ direction, because quasiparticle scattering between the projected gap nodes and/or the Fermi arc connecting the two gap nodes naturally occur at $E = 0$, and it will be quickly diminished when the energy moves away from zero.

Based on the above experimental arguments, the new wavevector $S_3$ we detect cannot represent a preexisting CDW nor a superconductivity induced CDW or PDW. However, it could represent quasiparticle scattering resulting from two superconducting gap nodes along the $\hat{a}$ direction.

## 11. Spherical Fermi surface of UTe$_2$

Measurements of UTe$_2$ Fermi surfaces have reached broad agreement regarding the existence of two cylindrical, quasi-2D bands with a hole-type band around the **X** point and an electron-type band around the **Y** point of the Brillouin zone (*45, 49*). However, these quasi-2D Fermi surfaces are associated with U-*6d* and Te-*5p* orbitals, despite consensus that heavy *f*-electron correlations and the Kondo effect should play an important role in the low-temperature electronic structure. How this heavy electron physics impacts the Fermi surface is a matter of ongoing debate. We therefore first review the experimental results.

Angle-resolved photoemission spectroscopy (ARPES) reports of the Fermi surface support the existence of a heavy U band centered on either the Γ (*44*) or Z point (*45*) of the Brillouin zone. This band is found to be approximately spherical and, from gradient analysis of the intensity map, has a *k*-space radius of ~0.2 Å$^{-1}$. Quantum oscillation experiments by Broyles *et al.* (*46*) found three low frequency, angle-independent peaks $F_α$, $F_β$, $F_γ$ indicative of a spherical pocket with radius ~0.2 Å$^{-1}$. However, Weinberger *et al.* (*50*) observe only one low frequency peak (206 T) and notably this peak does not exhibit the same angular dependence as those observed in Ref. 46. Instead of a closed Fermi surface Weinberger *et al.* propose that *f*-electron hybridization induces significant warping of the quasi-2D Fermi surfaces along the $k_z$ axis. The low frequency signal observed by Broyles *et al.* is then attributed to quantum interference effects between these hybridized bands (*50*).

Theoretical work has found that these interpretations of the Fermi surface are dependent on the degree of low-temperature hybridization. Calculations employing density functional theory (DFT) introduce an on-site Coulomb repulsion term U to describe *f*-electron correlations (*51*). Tuning this variable from 1 eV to 2 eV results in a Liftshitz transition of the Fermi surface at U ~ 1.6 eV. An intermediate value of U produces both the quasi-2D Fermi surface sheets and a pocket which encloses the Z point consistent with ARPES measurements. An intermediate value for U reflects more itinerant *f*-electrons expected for the Kondo effect at low temperature. Furthermore, both tight binding and DFT+DMFT (dynamical mean field theory) calculations of the Fermi surface reproduce this 3D Fermi surface component at low temperature and ambient pressure while also supporting $B_{3u}$ symmetry of the triplet order parameter (*52,53*).

Theoretical calculations therefore indicate that Kondo hybridization at low temperature increases the 3D character of the Fermi surface. This 3D character may manifest as a simple warping of the quasi-2D sheets or, for stronger hybridization, the Fermi surface may be forced to enclose the Z point generating a truly 3D Fermi surface. In any case, the presence of a Fano peak in differential conductance measurements as well as *c*-axis transport measurements highlight the role of Kondo coherence at low temperatures and suggest that the low-temperature electronic structure of $UTe_2$ must take these hybridization effects into account (*26,54*). Whether this hybridization is enough to enforce the presence of a closed, 3D Fermi surface is unresolved. Future experiments, including further low frequency quantum oscillation measurements or STM quasiparticle interference imaging, should contribute further to determining the true Fermi surface of $UTe_2$.

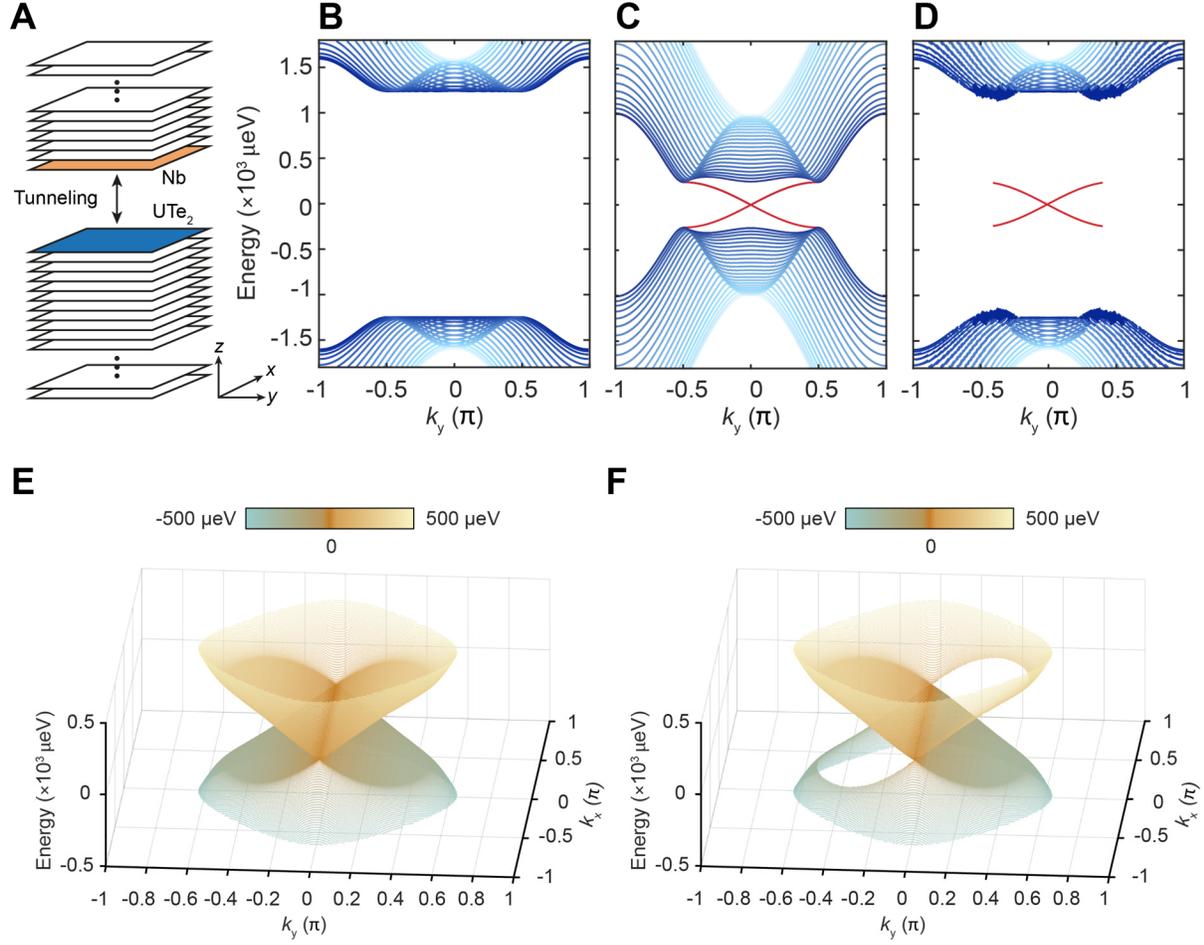

**Fig. S1. Slab calculations of SIP model.**
(**A**) Layered structure of SIP model stacked in the $z$ direction, and continuous in $x$, $y$ directions. (**B**) The surface band dispersion of Nb (indicated by the orange layer) when $M = 0$. In all the band dispersion calculations, the red dispersion lines denote the superconductive TSB. The shading of the blue dispersion lines highlight the low-energy band structure phenomena, and the excitations further from the Fermi level are less relevant to the tunneling process. (**C**) The surface band dispersion of UTe$_2$ (indicated by the blue layer) for non-chiral B$_{3u}$ pairing when $M = 0$. The TSB is presented in red to distinguish these states from those of the bulk. (**D**) The surface band dispersion of Nb on the orange layer when $M = 0.2$ meV. Bulk states of UTe$_2$ within the energy of the Nb superconducting gap are forbidden to tunnel into Nb. Only the TSB of UTe$_2$ can tunnel into the Nb surface, which contributes to the tunneling current near zero energy. It can lead to sharp zero bias peak in differential conductance measurements between UTe$_2$ and Nb. In the calculation, we plot the bands whose wavefunction weight exceeds a certain lower bound ($10^{-3}$ weight on the top surfaces of Nb and UTe$_2$, respectively). (**E**) 3D representation of the non-chiral TSB model used throughout. (**F**) 3D representation of the chiral TSB model used throughout.

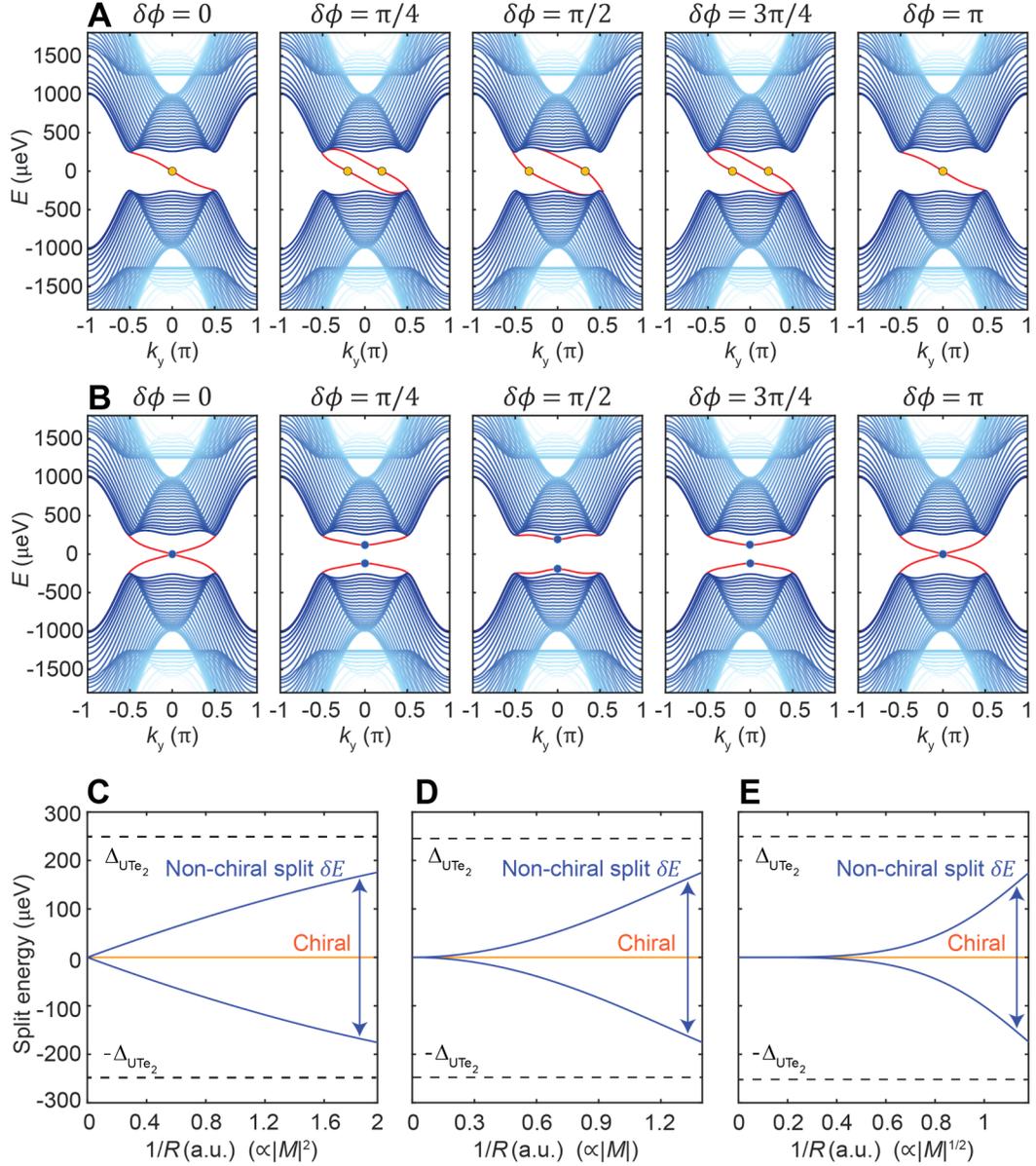

**Fig. S2. Evolution of in-gap states vs. the relative phase $\delta\phi$ between Nb and UTe$_2$.**
(**A**) Calculated quasiparticle bands within the SIP interface for chiral order parameter $A_u + iB_{3u}$. The chiral order parameter breaks time-reversal symmetry. The existence of its zero-energy states is therefore unaffected by tunneling to the *s*-wave Nb tip. (**B**) Calculated quasiparticle bands within the SIP interface for non-chiral order parameter $B_{3u}$. The tunneling matrix element $|M|$ is fixed while $\delta\phi$ evolves from 0 to $\pi$. The zero-energy states disappear due to broken time-reversal symmetry of the Nb-UTe$_2$ system at $\delta\phi = \pi/2$. (**C-E**) Energy splitting of zero-energy surface state derived from the model presented in Section 5. The relative phase $\delta\phi$ is kept fixed at $\pi/2$ while the tunneling matrix element $|M|$ is increased. We present three cases of split energy dependence on the tunneling matrix element $1/R \propto |M|^n$ with $n = 2, 1, 0.5$. $1/R \propto |M|$ in D is quantitively similar to the experimental data in Main Text Fig. 5D. Thus we choose to present D in Main Text Fig. 3D.

*39*

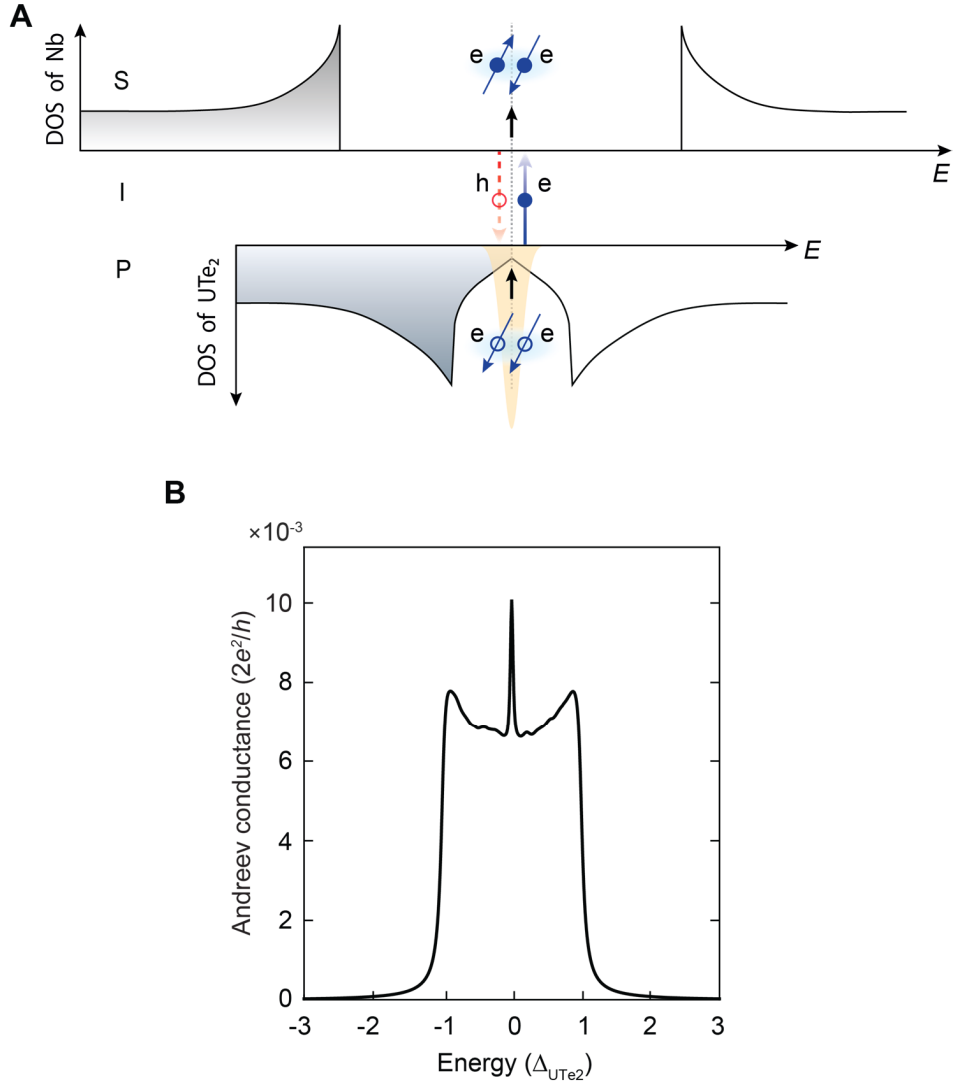

**Fig. S3. TSB generated Andreev conductance in the SIP model.**
(**A**) Schematic of the TSB generated Andreev tunneling to the *s*-wave electrode, through two quasiparticle transport process. (**B**) Calculated Andreev conductance $a(V)$ in the SIP model. Hence, the SIP model predicts a sharp peak in Andreev conductance surrounding zero-bias if the TSB is that of a *p*-wave, nodal, topological superconductor that mediates the *s*-wave to *p*-wave electronic transport processes. In this figure we have divided the total Andreev conductance by the number of transverse $k_\parallel$ channels to mimic the point tunneling of STM.

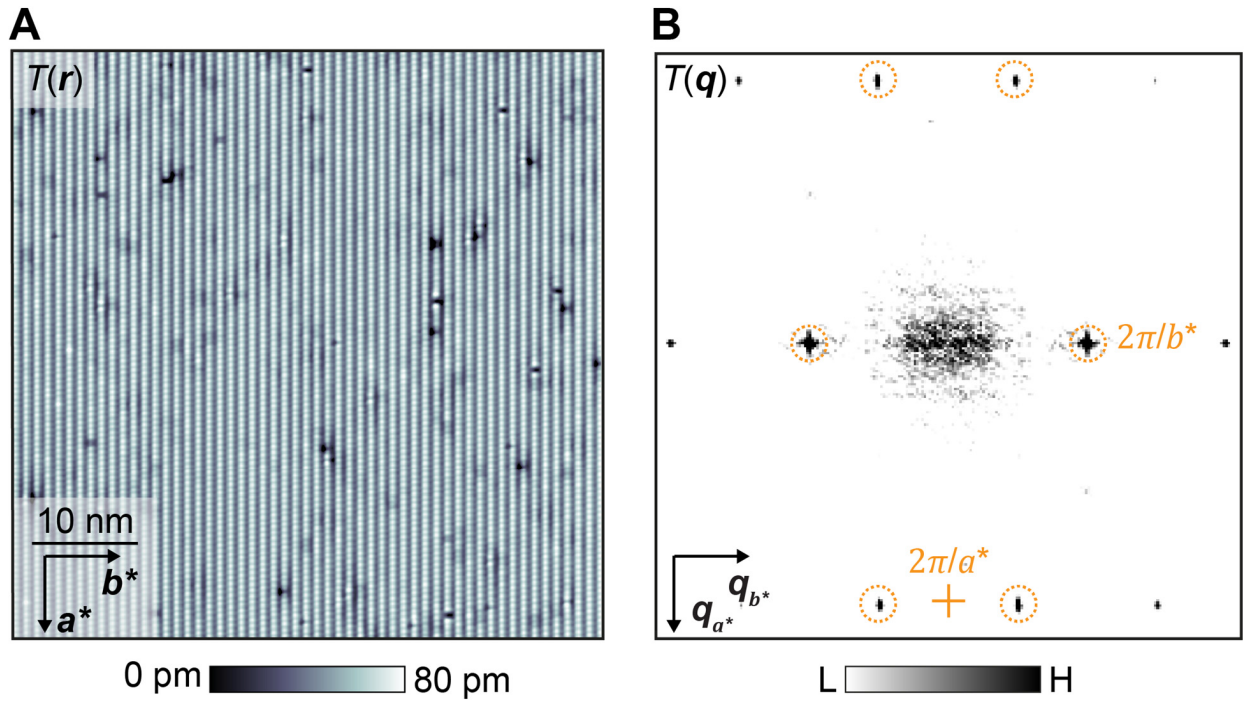

**Fig. S4. Topographic image measured by using superconducting tip.**
(**A**) Typical topographic image $T(r)$ of UTe$_2$ (0-11) surface measured with a superconducting STM tip. (**B**) Measured $T(q)$, the Fourier transform of $T(r)$, with the surface reciprocal-lattice points labelled as dashed circles.

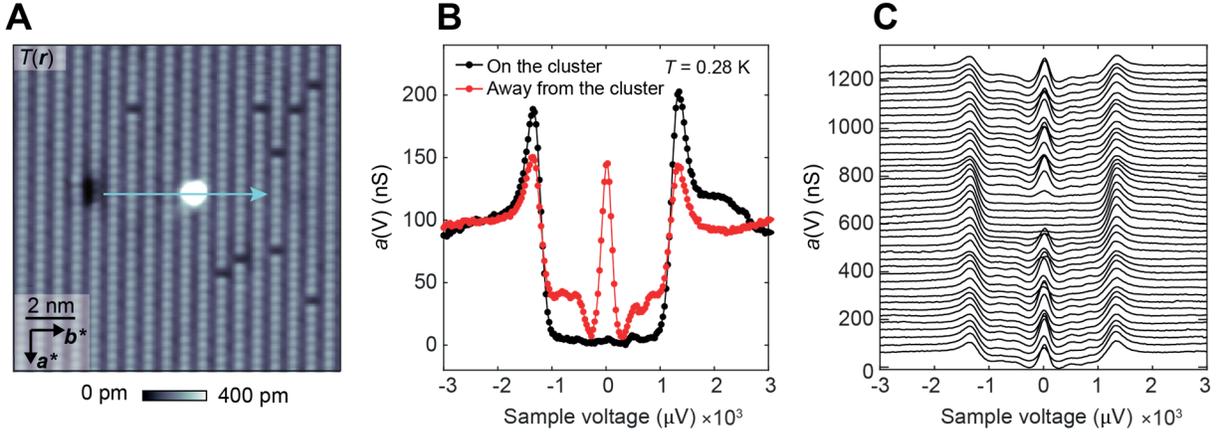

**Fig. S5. Linecut over impurity adatom cluster.**
(**A**) Topographic image of UTe$_2$ (0 –1 1) surface measured at $T$ = 280 mK. The high intensity near the center is a cluster of impurity atoms. (**B**) Differential conductance spectra recorded away from (red) and upon (black) the adatom cluster. Observed sub-gap features across the cleave surface of the UTe$_2$ falls to zero on the adatom cluster. (**C**) Evolution of the Andreev conductance across the impurity cluster measured along the blue arrow indicated in A. The conductance of sub-gap features collapses to zero as the STM tip measures across the impurity cluster.

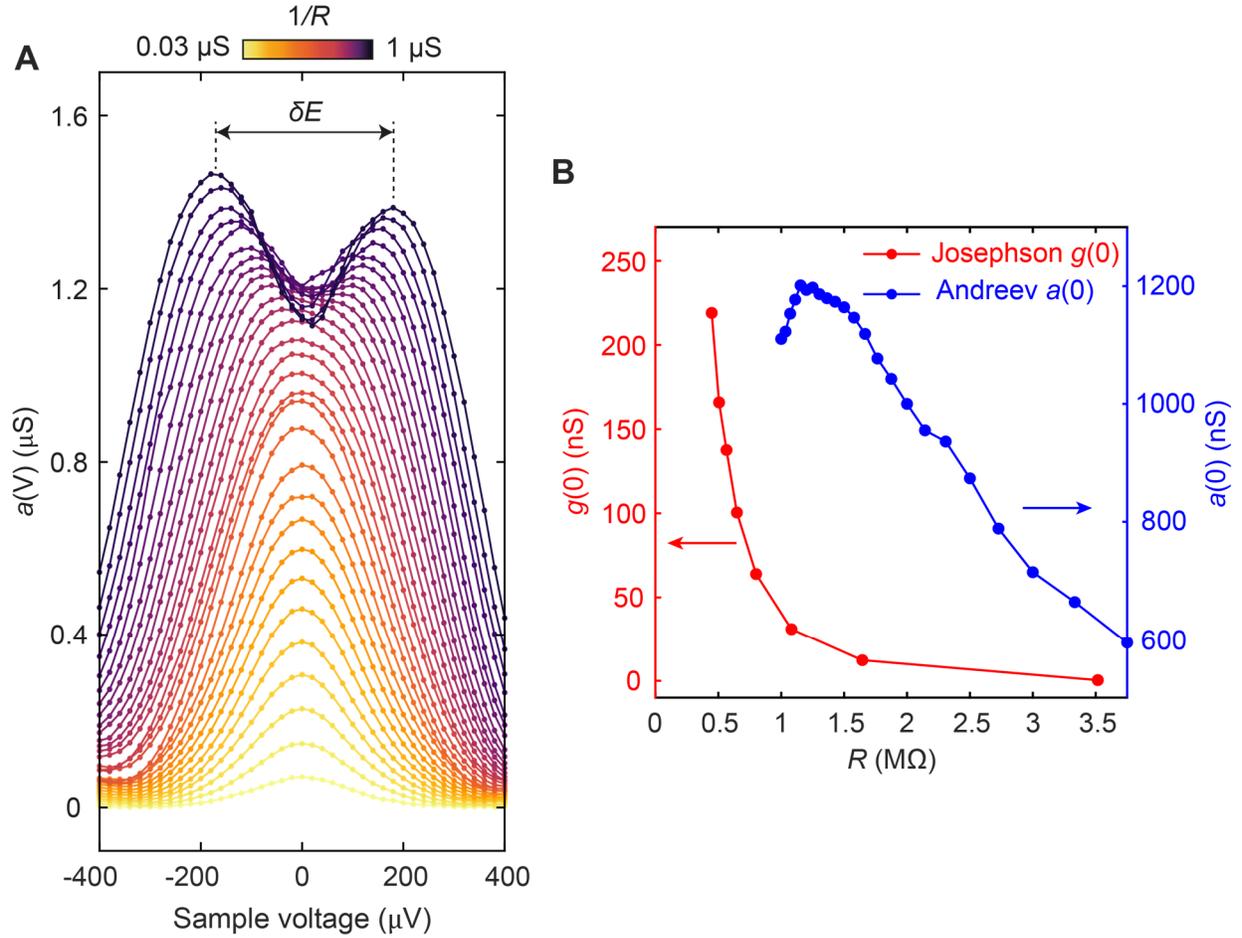

**Fig. S6. Intensity and evolution of $dI/dV|_{SIP}$ rules out Josephson currents.**
(**A**) Measured evolution of differential Andreev conductance ($a(V) \equiv dI/dV|_{SIP}$) spectra as a function of decreasing junction resistance $R$. (**B**) Comparison between the measured Andreev zero-bias conductance $a(0)$ of Nb/UTe$_2$ and Josephson zero-bias conductance $g(0)$ of Nb/NbSe$_2$ versus junction resistance $R$. The behaviour of the zero-bias conductance in the two effects are distinctly different and both the magnitude and $R$ dependence of $a(0)$ are strongly inconsistent with what is expected in the case of Josephson tunneling between Nb and UTe$_2$.

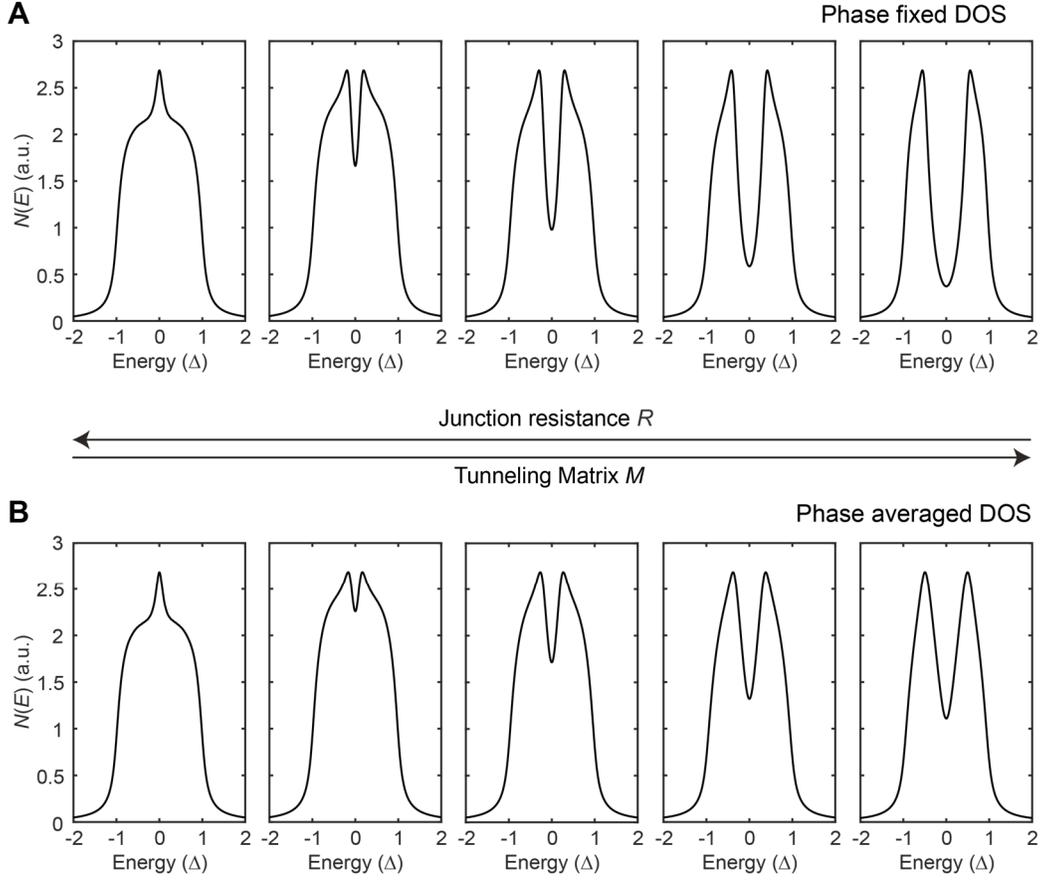

**Fig. S7. Phase fluctuation effect on tip-induced time-reversal symmetry breaking.**
(**A**) The calculated TSB density-of-states $N(E)$ when $\delta\phi = \frac{\pi}{2}$. (**B**) The calculated TSB density-of-states $N(E)$ when $\delta\phi$ is averaged with equal probability over the range $0 < \delta\phi < \pi$. The fact that the zero-bias peak splitting survives under these two extreme limits demonstrates that phase fluctuations will not destroy the signature of tip-induced time-reversal symmetry breaking in the SIP model.

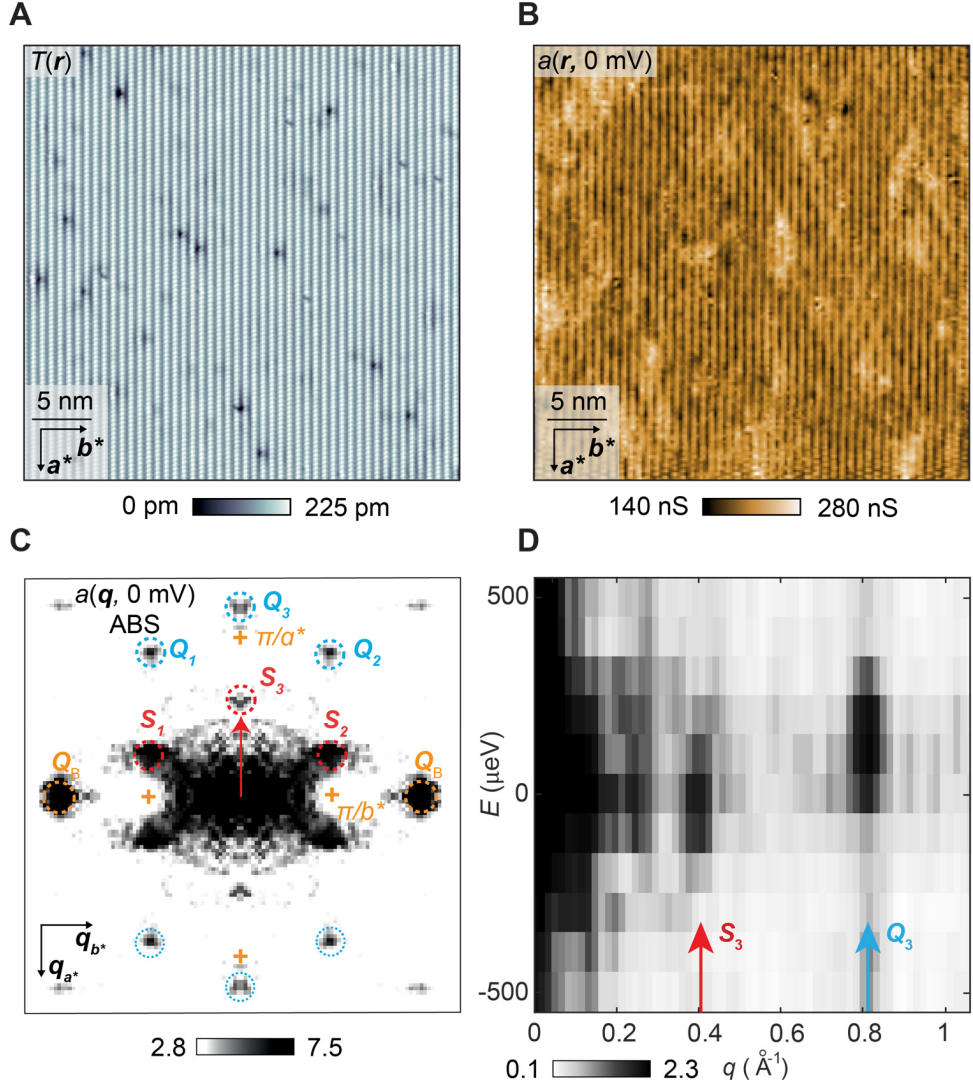

**Fig. S8. $S_3$ not consistent with superconductivity induced CDW.**
(**A**) Topograph of the (0 -1 1) cleave surface. (**B**) Andreev conductance $a(r, 0\,mV)$ map demonstrating the real space modulation of the zero-energy peak. It is measured at the same FOV as in A. (**C**) Power spectral density Fourier transform of B. Reciprocal lattice points are indicated by orange circles, CDW modulations are indicated by blue circles, and the three new scattering wavevectors $S_i$ ($i$ = 1,2,3) are labelled by red circles. (**D**) Linecut from (0, 0) to (0, 1) Å$^{-1}$ in C. The putative internodal scattering $S_3$ wavevector is indicated by a red arrow. The prevenient CDW modulation $Q_3$ is indicated by a blue arrow.

45